\documentclass[showpacs,aps,prb,twocolumn]{revtex4-1}  

\usepackage{amsmath,amssymb}
\usepackage{graphicx}
\PassOptionsToPackage{caption=false}{subfig} 
\usepackage{subfig}
\usepackage{soul} 
\usepackage{color}
\usepackage{hyperref}
\usepackage[all]{hypcap}
\usepackage{dsfont}

\usepackage{ulem}

\usepackage{grffile}

\usepackage{mathtools}

\def\mbs#1{\mbox{\tiny #1}}

\def\nnn{\nonumber \\}
\def\nn{\nonumber}

\def\op#1{\mathinner{\hat{#1}}}

\newcommand\dop[2]{\op{#1}_{#2}^{\dagger}}

\def\expval#1{\mathinner{\langle{#1}\rangle}}

\DeclareMathOperator{\e}{e}

\newcommand{\tr}[1]{\mbox{Tr}\left\{#1\right\}}

\newcommand{\trb}[1]{\mbox{Tr}_{\mbs{B}}\left\{#1\right\}}

\newcommand{\trs}[1]{\mbox{Tr}_{\mbs{S}}\left\{#1\right\}}

\newcommand{\comm}[2]{\left[#1,#2\right]}
\newcommand{\anticomm}[2]{\left\{#1,#2\right\}}

\def\calb{{\cal B}}

\def\calp{{\cal P}}

\def\call{{\cal L}}

\def\spup{\uparrow}
\def\spdo{\downarrow}

\def\bra#1{\mathinner{\left \langle {#1} \right \vert}}
\def\ket#1{\mathinner{\left \vert {#1} \right \rangle}}

\renewcommand{\imath}{\mathbf{i}}

\def\mat#1{\mathbf{#1}}

\newcommand{\parmat}[1]{\begin{pmatrix} #1 \end{pmatrix}}

\def\ip#1{\mbox{\boldmath $#1$}}

\newcommand{\nsigma}{\bar{\sigma}}

\begin{document}

\title{Semiclassical spin-spin dynamics and feedback control in transport through a quantum dot}

\author{Klemens Mosshammer}
\email{klemens@itp.tu-berlin.de}
\author{Tobias Brandes}
\affiliation{Institut f\"ur Theoretische Physik, Technische Universit\"at Berlin, Hardenbergstr. 36, D-10623 Berlin, Germany}

\date{\today}

\begin{abstract}
We present a theory of magnetotransport through an electronic orbital, where the electron spin interacts with a (sufficiently) large external spin via an exchange interaction. Using a semiclassical approximation, we derive a set of equations of motions for the electron density matrix and the mean value of the external spin that turns out to be highly nonlinear. The dissipation via the electronic leads is implemented in terms of a quantum master equation that is combined with the nonlinear terms of the spin-spin interaction. 
With an anisotropic exchange coupling a variety of dynamics is generated, such as self-sustained oscillations with parametric resonances or even chaotic behavior.
Within our theory we can integrate a Maxwell-demon-like closed-loop feedback scheme that is capable of transporting particles against an applied bias voltage and that can be used to implement a spin filter to generate spin-dependent oscillating currents of opposite directions.  
\end{abstract}

\pacs{
73.63.Kv,  
75.76.+j,  
85.75.-d,  
85.35.Gv   
}

\maketitle
\renewcommand{\theequation}{\arabic{equation}}
\renewcommand{\thefigure}{\arabic{figure}}
\renewcommand{\thetable}{\arabic{table}}


\section{Introduction}
In recent years research on single-electron transport through single 
molecules, quantum dots (QDs), or quantum wires \cite{Nazarov2009,Andergassen2010} has developed rapidly. Quantum transport is also a tool for studying spin states~\cite{Hanson2007} or coherent dynamics \cite{Shinkai2009} on a microscopic level. Recent research -- theoretically and experimentally -- provides some insights into level structures,~\cite{Reed1988,Reimann2002,Hanson2007} Coulomb- and spin blockade effects\cite{Ono2002}, phonon-emission \cite{Roulleau2011} and also full counting statistics of electron-tunneling processes~\cite{Bagrets2003,Fujisawa2006,Gustavsson2006a}.

Of particular interest for the electronic dynamics of QD electrons is the coupling to external degrees of freedom. Electrons tunneling to a QD device experience, for instance, a hyperfine and spin-orbit interaction with the nuclear spins of the host material. Transport experiments with QDs show that the hyperfine interaction can lift spin blockades and even induce self-sustained oscillations in currents~\cite{Ono2004} and that large Overhauser fields~\cite{Baugh2007} are experienced by the electrons.  The hyperfine interaction in QDs has also been studied theoretically in detail~\cite{Coish2004, Erlingsson2005,Inarrea2007,Rudner2011}.

Similar to a previous work \cite{Lopez-Monis2012} our model is inspired by experiments on the hyperfine interaction with QDs without transport where electronic spins in single QDs~\cite{Khaetskii2002,Reilly2008} or double QDs~\cite{Foletti2009,Bluhm2010} are considered in terms of spin relaxation and decoherence. There are also intriguing transport experiments where nonlinear behavior due to hyperfine interaction is induced: singlet-triplet state mixing in double QDs  that leads to transport bistabilities~\cite{Koppens2005}, the single-electron spin manipulation in a double QD~\cite{Koppens2006} or the lifting of spin blockades that leads to current fluctations driven by nuclear dynamics~\cite{Rudner2011}.

Exchange interactions that induce complex spin-spin dynamics also occur in molecular QDs. For the transport through molecular QDs~\cite{Galperin2008} two types of degrees of freedom are relevant: molecule vibrations~\cite{Hussein2010, Metelmann2011} or local magnetic moments in single molecular magnets~\cite{Heersche2006}, which establish the research field of molecular spintronics~\cite{Zutic2004,Fert2008,Bogani2008}. In the last couple of years a number of theoretical works have been done on models with a single orbital as current-carrying channel~\cite{Kiesslich2009,Bode2012,Baumgartel2011,Sothmann2010}.

The host material of QDs often contains a huge number of spins (nuclear or molecular) and can be described by one large effective spin. Interacting spins in transport models with isotropic exchange coupling have been studied recently, where a characteristic current induced switching of magnetic layers~\cite{Ralph2008,Brataas2012} or the external spin in QD setups~\cite{Bode2012,Mosshammer2012} was found as well as superradiant-like behavior in a single QD~\cite{Schuetz2012}. If the exchange coupling is anisotropic more involved nonlinear dynamics are spawned, that even contain chaos and which have been studied in detail for closed systems~\cite{Feingold1983,Magyari1987,Srivastava1990,Robb1998,Houle1999}. The nonlinear dynamics of anisotropically exchange coupled spins with connections to electronic reservoirs show intriguing features like self-sustained current oscillations, parametric oscillations, and chaotic dynamics. This has been addressed within different theoretical frameworks: In 
Refs.~\onlinecite{Lopez-Monis2012,Mosshammer2012} a quantum master equation with a classical large spin was used in an infinite-bias limit to study the transport characteristics of single- and double-QD setups. Metelmann \textit{et al.}\cite{Metelmann2012} derived the equations of motion by Keldysh-Green functions.

We are particularly interested in controlling the spin interactions by intervening in the transport process, i.e. applying a closed-loop feedback to our model. The goal of such an intervention is to prevent the system from running into chaotic regimes or fixed points. The feedback we think of is included on the level of the master equation and inspired by the notion of Maxwell's demon, capable of sorting particles by conditionally inserting/removing a wall. This mechanism ideally does not require work to insert or remove the wall, which modifies the entropy balance (i.e. the second law of thermodynamics) while not affecting the energy balance (first law)\cite{Esposito2012a}. A transport analog to the Maxwell demon is a device that is capable of generating electronic currents even against a bias voltage or thermal gradient by changing the energy barriers based solely on information about the current QD occupation. 
As an interesting application we show that our feedback scheme is capable of generating spin-currents of opposite directions.

Our Ref.~\onlinecite{Schaller2011} demonstrates the implementation of a demon-like feedback in a single-electron transistor and Ref.~\onlinecite{Strasberg2013} provides an insight on the thermodynamics of a physical 
implementation. Recent experiments also show that it is, in fact, possible to transform information about particles into free energy~\cite{Serreli2007,Toyabe2010,Berut2012}. For a transport setup the experimental difficulty is, clearly, to strongly modify the single-electron tunneling rates without changing the QD levels. However, investigations on quantum turnstile setups show that one can pump electrons by invoking a pump cycle based on the modulation of tunneling barriers by conventional electronics~\cite{Kouwenhoven1991,Blaauboer2005}. Within these cycles electronic levels are not changed.  
\\
\\
The remainder of this paper starts in Sec.~\ref{sec:model} with a detailed description of the model with the Hamiltonian (Sec.~\ref{sec:hamiltonian}), the master equation (Sec.~\ref{sec:me_transport}) and the introduction of the feedback mechanism (Sec.~\ref{sec:intro_feedback}).
The final equations of motion (EOMs) are presented in Sec.~\ref{sec:full_eom} and their resulting dynamics are discussed in Sec.~\ref{sec:results}. First we discuss the results for the transport without spin-spin interactions (Sec.~\ref{sec:results_noninteracting}).  
We proceed with discussing the results of solutions for the full system for infinite-bias voltages (Sec.~\ref{sec:infinite_bias}) and the results for the finite-bias regime are provided in Secs.~\ref{sec:finite_bias} and \ref{sec:discussion_trajectory}. Finally, we conclude in Sec.~\ref{sec:conclusions}.


\section{\label{sec:model}Model}
\subsection{\label{sec:hamiltonian}Hamiltonian}
\begin{figure}[htb]
\includegraphics[width=0.4\textwidth]{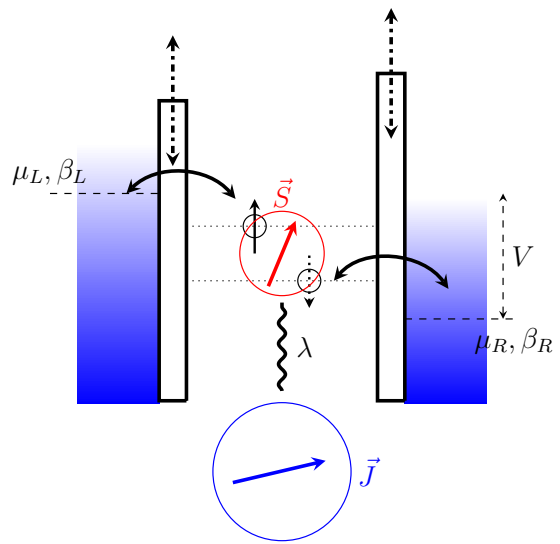}
\caption{(Color online) Setup of the investigated system. An electronic spin $\vec{\op{S}}$ (red) in a single quantum dot (SQD) is coupled to an external spin $\vec{\op{J}}$ (blue) via an exchange interaction $\lambda$ (wiggly line). The SQD level is split up by an external magnetic field $\vec{B}$ in the $z$ direction. Due to coupling to the leads $l$ (characterized by inverse temperatures $\beta_l$ and chemical potentials $\mu_l$) electronic transport is taking place on and off the SQD (solid arrows). The feedback mechanism, introduced in Sec.~\ref{sec:intro_feedback}, instantaneously modifies the tunneling barriers conditioned on the system states (dash-dotted arrows).  \label{fig:setup}  
}
\end{figure}

We consider a system of a single quantum dot (SQD) with one orbital level that is subject to an external magnetic field $\vec{B}$ in the $z$ direction which splits the QD level (see Fig.~\ref{fig:setup}). The SQD is coupled to electronic leads and without any further interaction the coupling leads to the formation of two distinct spin-dependent current channels, since the spin of the tunneling electrons is assumed to be invariable while tunneling.

Moreover, the model consists of a large spin (LS) $\vec{\op{J}}$ with length $j$ given by $\vec{\op{J}}^2\ket{m,j} = j(j+1)\ket{m,j}$, the $z$ component of which couples to the magnetic field as well and which is exchange coupled with the electron spin. For simplicity, we include the $g$ factors of the electronic spin and the LS and the Bohr magneton in our definition of $B$ and assume the $g$ factors to be the same for electronic and LS. 

The full Hamiltonian reads as follows~\cite{Lopez-Monis2012,Mosshammer2012,Metelmann2012}:
\begin{align}
\label{eq:hamiltonian}
 \op{H} &= \op{H}_{\mbs{SQD}} + \op{H}_{\mbs{LS}} + \op{H}_{\mbs{int}} + \op{H}_{\mbs{leads}} + \op{H}_{\mbs{T}}\,, \nnn
 \op{H}_{\mbs{SQD}} &= \sum_{\sigma = \spup,\spdo} \varepsilon \dop{d}{\sigma} \op{d}_\sigma + B \op{S}_z\,, \nnn
 \op{H}_{\mbs{LS}} &= B \op{J}_z\,,\quad \op{H}_{\mbs{int}} = \sum_{i=x,y,z} \lambda_i \op{S}_i \op{J}_i  \,, \nnn
 \op{H}_{\mbs{leads}} &= \sum_{kl\sigma} \varepsilon_{kl\sigma} \dop{c}{kl\sigma} \op{c}_{kl\sigma} \quad \quad (l = L/R) \,,\\
 \op{H}_{\mbs{T}} &= \sum_{kl\sigma} \left(\gamma_{kl\sigma} \dop{c}{kl\sigma} \op{d}_{\sigma} + \mbox{h.c.} \right)\,. \nn
\end{align}
The operators $\dop{d}{\sigma}(\op{d}_{\sigma})$ describe the creation (annihilation) of an electron with spin $\sigma = \spup, \spdo$ on the dot, $\op{n}_\sigma$ is the related occupation number operator, and $\op{S}_{i}$ are the components of the electronic spin operators in second quantization and can be written in terms of the creation/annihilation operators
\begin{align}
 \label{eq:components_of_electronic_spin}
 \op{S}_x &= \frac{1}{2} \Big(\dop{d}{\spup} \op{d}_{\spdo} + \dop{d}{\spdo} \op{d}_{\spup} \Big) = \frac{1}{2} \Big(\op{S}_+  + \op{S}_-\Big) \,, \nnn
 \op{S}_y &= \frac{1}{2 \imath} \Big(\dop{d}{\spup} \op{d}_{\spdo} - \dop{d}{\spdo} \op{d}_{\spup} \Big) = \frac{1}{2 \imath } \Big(\op{S}_+ - \op{S}_-\Big)\,, \\
 \op{S}_z &= \frac{1}{2} \Big(\op{n}_{\spup} -  \op{n}_{\spdo} \Big)\nn \,,
\end{align}
with the usual commutation relations for angular momentum operators $(i,j,k=x,y,z, \hbar=1)$,
\begin{align}
\label{eq:angular_momentum_commutator}
 \comm{\op{S}_i}{\op{S}_j} &= \imath \sum_k \varepsilon_{i j k} \op{S}_k \,,
\end{align}
where $\varepsilon_{ijk}$ is the Levi-Civita-symbol. 

The electronic leads are assumed to be noninteracting. Electrons of momentum $k$ and spin $\sigma$ in the $l$-th lead are created(annihilated) by the corresponding operators $\dop{c}{kl\sigma}(\op{c}_{kl\sigma})$. The transitions between a state in the leads and the electronic levels are described by $\op{H}_{\mbs{T}}$, and the tunneling amplitudes for these transitions are $\gamma_{kl\sigma}$.

\subsection{\label{sec:me_transport}Master equation for exchange coupling-assisted transport}
The microscopic dynamics of the system described by the Hamiltonian \eqref{eq:hamiltonian} is involved. In particular, we are interested in the dynamics of single-particle observables such as components of the average electron and LS and we consider the reduced density matrix of electrons dwelling in the dots.

The system's dynamics are governed by different time scales. 
First of all, we assume that the electronic leads are in thermal equilibrium and we have the time scale $\tau_c$ on which the bath correlations decay during a tunneling event. The second time scale is set by the electron tunneling, which we assume is much faster than the precession of the LS; that implies that electron spin fluctuations do not affect the LS dynamics and vice versa, which is reflected by a mean-field approximation.
Another time scale is set by the instantaneous feedback mechanism and depends only on the occupation of the SQD.

Furthermore, under the assumption that the coupling between electronic leads and the SQD is weak, we can use the Born-Markov (BM) approximation.\cite{Breuer2002,Schaller2009} The resulting BM master equation \eqref{eq:me_with_dissipators_full_form} governs the evolution of SQD density operator $\op{\rho}_{\mbs{S}}$. The Liouvillian superoperator $\op{\call}$ derived in Appendix \ref{sec:derivation_master_equation} is not only parametrized by the Markovian system-bath tunneling rates $\Gamma_{l\sigma}(\omega) = 2\pi \sum_k |\gamma_{kl\sigma}|^2 \delta(\varepsilon_{kl\sigma} - \omega)$ and the lead Fermi functions $f_{l} (\omega) = [\e^{\beta(\omega - \mu_l)} + 1]^{-1}$ but also by the (slow) dynamics of the LS $\vec{\op{J}}$ and in particular by its interaction with the electrons dwelling in the SQD. Within the derivation a mean-field approach, that neglects the (fast) fluctuations of the LS, has been made, as carried out in \ref{sec:mean-field_approximation}. 
Therefore, the influence of the LS is reflected by the effective Zeeman splitting $\varepsilon_z = B + \lambda_z \expval{\op{J}_z}_t$ that has a contribution coming from the LS's polarization. On the other hand the exchange coupling induces flips of the electronic spins driven by the parameter $\Lambda = \frac{\lambda_x}{2} \expval{\op{J}_x}_t + \imath \frac{\lambda_y}{2} \expval{\op{J}_y}_t$.

\subsection{\label{sec:intro_feedback}Introducing feedback}
For small bias voltages across the SQD a device that yields information about its instantaneous occupation, such as an quantum point contact (QPC), cannot resolve to which of the attached electron reservoirs an electron has tunneled. Therefore, the simplest feedback mechanisms can only be conditioned on the occupation of the SQD itself or whether an electron has tunneled in or out, without knowing the direction. Recent experiments show that the random telegraph signals of QPCs contains information about the spin state of electrons in monitored QDs \cite{House2013}.
Now the Maxwell demon feedback is implemented as suggested in Ref.~\onlinecite{Schaller2011}. We apply different Liouvillians conditioned on whether the SQD is empty ($E$), populated in the $\spup$-state or populated in the $\spdo$-state; i.e. we construct new rate matrices by taking the rate matrix \eqref{eq:full_Liouvillian} and multiplying it by the dimensionless feedback parameters $\delta_{l \nu} \in \mathbb{R}; \nu \in \{E,\spup,\spdo\}$, which encode the modification of the tunneling rates (with $\delta_{l \nu} = 0$ recovering the case without feedback),

\begin{align}
 \call_\nu & \equiv \sum_{l=L/R} \e^{\delta_{l \nu}} \call^{(l)}\,.
\end{align}

We construct the effective feedback generator by projecting on the empty and filled dot states and on the two interesting coherences,

\begin{align}
 \label{eq:fb_liouvillian}
 \call_{\mbs{fb}} &= \call_{E}
\begin{pmatrix}
 1 & 0 & 0 & 0 & 0\\
 0 & 0 & 0 & 0 & 0\\
 0 & 0 & 0 & 0 & 0\\
 0 & 0 & 0 & 0 & 0\\
 0 & 0 & 0 & 0 & 0
\end{pmatrix} + \call_{\spdo}
\begin{pmatrix}
 0 & 0 & 0 & 0 & 0\\
 0 & 1 & 0 & 0 & 0\\
 0 & 0 & 0 & 0 & 0\\
 0 & 0 & 0 & 0 & 0\\
 0 & 0 & 0 & 0 & 0
\end{pmatrix} \nnn
& + \call_{\spup}
\begin{pmatrix}
 0 & 0 & 0 & 0 & 0\\
 0 & 0 & 0 & 0 & 0\\
 0 & 0 & 1 & 0 & 0\\
 0 & 0 & 0 & 0 & 0\\
 0 & 0 & 0 & 0 & 0
\end{pmatrix} + \call_{\mbs{nofb}} 
\begin{pmatrix}
 0 & 0 & 0 & 0 & 0\\
 0 & 0 & 0 & 0 & 0\\
 0 & 0 & 0 & 0 & 0\\
 0 & 0 & 0 & 1 & 0\\
 0 & 0 & 0 & 0 & 1
\end{pmatrix}
\,,
\end{align}
where $\call_{\mbs{nofb}} = \call$ is the non-modified Liouvillian, \eqref{eq:full_Liouvillian}.

This feedback scheme has been named ``Maxwell demon feedback'' by Schaller and Esposito in Refs.~\onlinecite{Schaller2011,Esposito2012a}, as one can think of a Maxwell demon that is able to instantaneously alter the tunneling amplitudes for the hopping on and off the SQD only based on the information about the current state of the system.

\subsection{\label{sec:full_eom}Final EOM of interacting spin system with demon like feedback}
The final equations that are subject of our investigation read
\begin{align}
 \frac{\partial}{\partial t} \vec{\rho} &= \call_{\mbs{fb}} \vec{\rho}\,. \label{eq:me_vector_form}
\end{align}

where the vector $\vec{\rho}$ comprises the relevant observables we need to describe the nonlinear spin-spin dynamics under feedback
\begin{align}
 \vec{\rho} &= \left[\expval{\rho_{00}}, \expval{\op{n}_{\spdo}}, \expval{\op{n}_{\spup}}, \expval{\op{S}_+},  \expval{\op{S}_-} \right]^T\,. 
\end{align}

We now can reduce the amount of equations by eliminating the equation for $d/dt \expval{\rho_{00}}$ by using $\tr{\rho} = 1$, which is possible because the transport rate matrix $\call_{\mbs{fb}}$ has rank 4. We obtain with \eqref{eq:components_of_electronic_spin} (for the sake of readability we omit the time dependences, here)

\begin{align}
\label{eq:eom_sqd}
\frac{d}{dt} \expval{\op{S}_x} &= +\lambda_y \expval{\op{J}_y} \expval{\op{S}_z} - \varepsilon_z \expval{\op{S}_y} - \sum_{l \sigma} \frac{\Gamma_{l\sigma}}{2} \overline{f_{l\sigma}} \expval{\op{S}_x} \,, \nnn
\frac{d}{dt} \expval{\op{S}_y} &= -\lambda_x \expval{\op{J}_x} \expval{\op{S}_z} + \varepsilon_z \expval{\op{S}_x} - \sum_{l \sigma} \frac{\Gamma_{l\sigma}}{2} \overline{f_{l\sigma}} \expval{\op{S}_y} \,, \nnn
\frac{d}{dt} \expval{\op{S}_z} &= \sum_{l} \Bigg\{\e^{\delta_{l E}} \left[\frac{\Gamma_{l\spup}}{2} f_{l \spup} - \frac{\Gamma_{l\spdo}}{2} f_{l \spdo} \right] \left(1 - \expval{\op{n}_\spup} - \expval{\op{n}_\spdo} \right)  \nnn
& \quad - \e^{\delta_{l \spup}} \frac{\Gamma_{l\spup}}{2} \overline{f_{l\spup}}\expval{\op{n}_\spup} + \e^{\delta_{l \spdo}} \frac{\Gamma_{l\spdo}}{2} \overline{f_{l\spdo}} \expval{\op{n}_\spdo} \Bigg\}  \nnn
& \quad + \lambda_x \expval{\op{J}_x} \expval{\op{S}_y} -\lambda_y \expval{\op{J}_y} \expval{\op{S}_x} \,, \\
\frac{d}{dt} \expval{\op{n}_\sigma} &= \sum_l \e^{\delta_{l E}} \Gamma_{l\sigma} f_{l\sigma} (1 - \expval{\op{n}_\spup} - \expval{\op{n}_\spdo}) \nnn
 & \quad - \sum_l \e^{\delta_{l \sigma}} \Gamma_{l\sigma} \overline{f_{l\sigma}} \expval{\op{n}_\sigma} \nnn
 & \quad + \left(\lambda_x \expval{\op{J_x}} \expval{\op{S}_y} - \lambda_y \expval{\op{J}_y} \expval{\op{S}_x}\right) \left(\delta_{\sigma\spup}-\delta_{\sigma\spdo}\right) \nonumber\,.
\end{align}

These equations are highly nonlinear and we need to generate EOM for the components of the LS to complete the set of equations.

The LS components obey the commutation relations $\comm{\op{J}_i}{\op{J}_j} = \imath \sum_k \varepsilon_{ijk} \op{J}_k$.

In our derivation of the BM master equation, cf. App.\ref{sec:derivation_master_equation} we implemented a mean-field approximation, which implies that the LS is not decaying due to the electronic leads; i.e. its length $j$ is conserved, and $\vec{\op{J}}$ is essentially treated as a classical object. Therefore, we can use the mean-field interaction Hamiltonian \eqref{eq:mean-field_interaction_hamiltonian} to calculate the Ehrenfest EOM $\frac{d}{d t} \expval{\op{J}_i} = -\imath \expval{\comm{\op{J}_i}{\op{H}_{\mbs{LS}} + \op{H}_{\mbs{int}}^{\mbs{MF}}}} + \expval{\frac{\partial \op{J}_i}{\partial t}}$. 
We obtain the EOM of the expectation values of the LS's degrees of freedom $\expval{\op{J}_i}$,
\begin{align}
\label{eq:eom_largespin_expanded}
\frac{d}{dt} \expval{\op{J}_x} 
& = + \lambda_y \expval{\op{S}_y} \expval{\op{J}_z} - \left(B + \lambda_z \expval{\op{S}_z} \right) \expval{\op{J}_y} \,, \nnn
\frac{d}{dt} \expval{\op{J}_y} 
& = - \lambda_x \expval{\op{S}_x} \expval{\op{J}_z} + \left(B + \lambda_z \expval{\op{S}_z}\right) \expval{\op{J}_x} \,, \\
\frac{d}{dt} \expval{\op{J}_z} 
& = + \lambda_x \expval{\op{S}_x} \expval{\op{J}_y} - \lambda_y \expval{\op{S}_y} \expval{\op{J}_x} \nonumber \,.
\end{align}
These equations have the form of Hasegawa-Bloch equations~\cite{Bloch1946,Hasegawa1986} completed by the electronic backaction.

\section{\label{sec:results}Analysis and Numerical results}

\subsection{\label{sec:results_noninteracting}System without exchange-interaction ($\lambda_i = 0$)}
If the exchange-interaction is ineffective we do not need to consider any coherences as the $\spup$- and $\spdo$-transport channels do not mix. With the new vector of probabilities $p_\nu$ to find the system in states $\nu = \{E,\spdo,\spup\}$, $\vec{\rho}_{\mbs{noint}}  = \left[p_E,p_\spdo,p_\spup \right]^T = \left[\expval{\rho_{00}}, \expval{\op{n}_{\spdo}}, \expval{\op{n}_{\spup}} \right]^T$ Eq. \eqref{eq:me_vector_form} reduces to $\frac{\partial}{\partial t} \vec{\rho}_{\mbs{noint}} = \call_{\mbs{fb}}^{(\mbs{noint})} \vec{\rho}$, where the rate matrix is merely the $3\times3$ submatrix of \eqref{eq:fb_liouvillian} with respect to the occupations, i.e. 
\begin{align}
 \call_{\mbs{fb}}^{(\mbs{noint})} &= \sum_l 
\parmat{
-\sum_\sigma W_{\sigma E}^{(l)} & W_{E \spdo}^{(l)} & W_{E \spup}^{(l)} \\
 W_{\spdo E}^{(l)} & - W_{E \spdo}^{(l)} & 0 \\
 W_{\spup E}^{(l)} & 0 & - W_{E \spup}^{(l)}
 } \,.
\end{align}

Here the rates for transitions between unoccupied SQD ("$E$``) and spin-$\sigma$ states induced by the coupling to lead $l$ read
\begin{align}
 W_{\sigma E}^{(l)} &= \e^{\delta_{lE}} \Gamma_{l \sigma}  f_{l}(\varepsilon_\sigma)\,, \quad W_{E \sigma}^{(l)} = \e^{\delta_{l\sigma}} \Gamma_{l \sigma}  \overline{f_{l\sigma}(\varepsilon_\sigma)}\,.
\end{align}
Along the lines of Ref.~\onlinecite{Esposito2012a} it can be shown straightforwardly, that the local detailed balance condition of our device is modified once the feedback is applied; i.e., constants $\Delta_{l\sigma} \equiv \delta_{l\sigma} - \delta_{lE}$ are nonzero,
\begin{align}
 \ln \frac{W_{\sigma E}^{(l)}}{W_{E \sigma}^{(l)}} &= -\beta_l (\varepsilon_\sigma - \mu_l) - \Delta_{l\sigma}\,.
\end{align}

A detailed discussion of the stochastic thermodynamics of the Maxwell demon feedback has been provided in Refs.~\onlinecite{Esposito2012a,Strasberg2013}. The crucial point is that the feedback mechanism does not affect the energy and matter balances of the system but does alter the entropy balance; i.e, besides the matter and energy currents $I_{\mbs{M}}$ and $I_{\mbs{E}}$, an information current $I_{\mbs{F}}$ is introduced, that changes the second law of the system.

The currents from lead $l$ read
\begin{align}
 I_{\mbs{E}}^{(l)} &= \sum_{\nu \neq \nu'} W_{\nu \nu'}^{(l)} p_{\nu'} (\varepsilon_\nu - \varepsilon_{\nu'}) \nnn
 I_{\mbs{M}}^{(l)} &= \sum_{\nu \neq \nu'} W_{\nu \nu'}^{(l)} p_{\nu'} (N_\nu - N_{\nu'}) \nnn
 I_{\mbs{F}}^{(l)} &= k_B \sum_{\nu \neq \nu'} W_{\nu \nu'}^{(l)} p_{\nu'} (\delta_{l \nu} - \delta_{l \nu'})\,,
\end{align}
 where $\varepsilon_{\nu}$, $N_{\nu}$ are the respective energies and particle numbers. Since we restrict the number of particles on the SQD to 1, the matter and energy currents are proportional to each other.
 The electron tunneling is spin preserving and, thus, we can treat the $\spup/\spdo$ channels separately and find (note that, due to particle conservation, $I_{\mbs{M}\sigma} = I_{\mbs{M}\sigma}^{(L)} = - I_{\mbs{M}\sigma}^{(R)}$)
\begin{align}
 I_{\mbs{M}\sigma} = W_{\sigma E}^{(L)} p_E - W_{E \sigma}^{(L)} p_\sigma &= I_{\sigma} = I_{\mbs{E}\sigma}/\varepsilon_\sigma \,.
\end{align}
If the system is in its stationary state the spin-$\sigma$ currents read
\begin{align}
\label{eq:spincurrent_no_large_spin}
 I_{\sigma} &= \frac{ \left(\e^{(\delta_{LE} + \delta_{R\sigma})} f_{L\sigma} \overline{f_{R\sigma}} -
      \e^{(\delta_{L\sigma} + \delta_{RE})} \overline{f_{L\sigma}} f_{R\sigma}\right) \Gamma_{L\sigma} \Gamma_{R\sigma}}
{\sum_l \left(\e^{\delta_{lE}} f_{l\sigma} + \e^{\delta_{l\sigma}} \overline{f_{l\sigma}} \right)\Gamma_{l\sigma} +C_\sigma} \,, \nnn
%
%
C_\sigma &= \frac{\left(\sum_l \e^{\delta_{lE}} f_{l\nsigma} \Gamma_{l\nsigma} \right) \left(\sum_{l} \e^{\delta_{l\sigma}} \overline{f_{l\sigma}} \Gamma_{l\sigma}\right)}
{\left(\sum_l \e^{\delta_{l\nsigma}} \overline{f_{l\nsigma}} \Gamma_{l\nsigma}\right)} \,,
\end{align}
where negative currents are effective currents from the right to the left lead and positive ones vice versa. The information current, on the other hand, evaluates to 
\begin{align}
 I_{\mbs{F}} = k_B \sum_{\sigma} (\Delta_{L \sigma} - \Delta_{R \sigma}) I_{\sigma}\,.
\end{align}

The system entropy is given in terms of the Shannon entropy,
\begin{align}
 S = -k_B \sum_\nu p_\nu \ln p_\nu \,.
\end{align}

Consequently, $\dot{S}$ is the entropy change in the system and after some algebra we can define the total entropy production~\cite{Esposito2012a},
\begin{align}
 \dot{S}_{\mbs{i}} &= \dot{S} - \sum_l \frac{\dot{Q}^{(l)}}{T_l} + I_{\mbs{F}}\,,
\end{align}
which is non-negative definite and where $\dot{Q}^{(l)} = I_{\mbs{E}}^{(l)} - \mu_l I_{\mbs{M}}^{(l)}$ is the heat flow with the $l$-th reservoir. 

In the absence of feedback ($I_{\mbs{F}} = 0$) the entropy production is just the sum of entropy change in the system and the reservoirs. The non-negativity of $\dot{S}_{\mbs{i}}$ implies that $\dot{S} \ge \sum_l \dot{Q}^{(l)}/T_l$, which is the second law of thermodynamics. The presence of feedback may change this, depending on the sign of $I_{\mbs{F}}$. The entropy change provided by the feedback mechanism adds to the total entropy. Once the system reaches its steady state the system entropy remains constant ($\dot{S}=0$) and the entropy production becomes
\begin{align}
 \dot{S}_{\mbs{i}} &= \left(\frac{1}{T_R}-\frac{1}{T_L}\right) I_{\mbs{E}} -\left(\frac{\mu_R}{T_R}-\frac{\mu_L}{T_L}\right) I_{\mbs{M}} + I_{\mbs{F}} \ge 0.  
\end{align}
If the leads are held at the same temperature $T_R = T_L = T$ and a chemical potential gradient is established ($\mu_R - \mu_L \ge 0$), the extracted power is $P=\left(\mu_R-\mu_L\right) I_{\mbs{M}}$. In the absence of feedback $I_{\mbs{F}}$ the matter flux can only flow with the gradient ($I_{\mbs{M}} \le 0$). If the feedback mechanism is effective and if the feedback current is sufficiently positive, particles can be transferred against the chemical potential gradient ($I_{\mbs{M}} \ge 0$).


If the device can discriminate spin directions of electrons we can implement the following feedback schemes. 

\begin{figure}[t!]
\label{fig:demon_currents}
\includegraphics[width=0.49\textwidth,keepaspectratio=true]{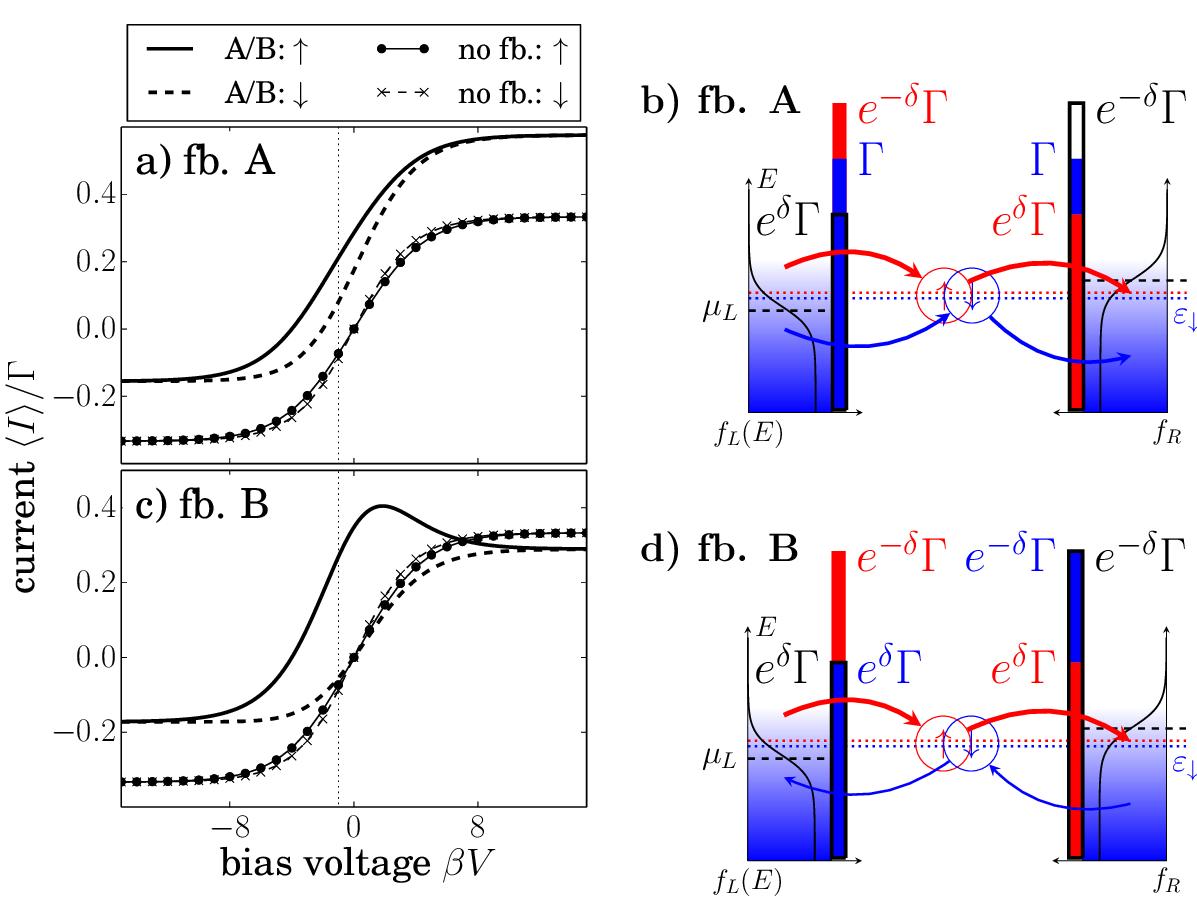}
 \caption{(Color online) Plots a) and c) show the spin-dependent currents as function of the applied bias voltage $V$ between the left and right transport leads, plotted for different feedback schemes and compared to the nonfeedback case. $\spup$($\spdo$) currents are depicted as solid (dashed) lines. Scheme B can act as a spin filter, as depicted in c), where only $\spup$-electrons are transported against the bias. For scheme A, shown in a), all electrons are transported against the bias, but $\spup$ currents are preferred. Plots b) and d) show a sketch of the SQD with the respective tunneling rates if the system is empty(black) or occupied by an $\spup$ (red) or $\spdo$ (blue) electron. Thickness and directions of the red/blue tunneling arrows show which net currents are reached. The setup is sketched for fixed bias voltages, $V/\lambda = 1$, shown by the dotted vertical lines in plots a) and c).}
\end{figure}

\textit{Feedback schemes.} In this paper we investigate the effect of two feedback parameter choices. First is scheme A with parameters chosen as $\delta_{L\spup} = \delta_{RE} = -\delta, \delta_{R\spup} = \delta_{LE} = \delta , \delta_{L\spdo} = \delta_{R\spdo} = 0$, where $\delta$ is positive.
In scheme A currents of both $\spup$ and $\spdo$ electrons can be transported against a chemical bias, but the $\spup$ currents are preferred. 
The second scheme, B, implements a spin filter, where $\spup$-currents are pumped against the bias and $\spdo$ currents are transported with the bias, by applying the parameters $\delta_{L\spup} = \delta_{RE} = \delta_{R\spdo} = -\delta$ and $\delta_{L\spdo} = \delta_{LE} = \delta_{R\spup} = \delta$.
The physical effect of the two schemes can be seen best in Fig. \ref{fig:demon_currents}, where in subplots a) and c) the current-bias voltage characteristics are plotted and in panels b) and d) the tunneling processes are sketched. Without feedback, all tunneling rates are equal, $\Gamma_{l\sigma} = \Gamma$. Due to the feedback the rates are according to the schemes. The averaged tunneling is depicted by the red and blue arrows, the thickness and direction of which indicate the direction and strength, respectively, of the resulting spin-$\sigma$ currents (red arrows show $\spup$ currents, while blue ones show $\spdo$ currents).

The spin filter effects can be used to generate oscillating currents with or against the bias depending on the electron spin in the interacting setup, as shown in Sec.~\ref{sec:discussion_trajectory}.

\subsection{\label{sec:closed_system_dynamics}Closed system dynamics ($\gamma_{kl\sigma} = 0$)}
In a number of previous works \cite{Magyari1987,Srivastava1990,Robb1998,Houle1999} the anisotropic exchange interaction between two classical spins in a closed system and under the presence of an external magnetic field was a subject of investigation. The exchange interaction between the spins induces nonlinear dynamics of the two spins.
Magyari \textit{et.al.}~\cite{Magyari1987} found that for arbitrary coupling $\lambda_i$ and vanishing magnetic fields the system is completely integrable. For an anisotropic coupling, however, the system becomes nonintegrable for finite magnetic fields, which can even lead to chaotic classical dynamics. 
One finds as well that in the limit $r=j/s \gg 1$, where $s,j$ are the spins' magnitudes, the larger spin $\vec{\op{J}}$ will act as an external ``driving'' for the smaller spin $\vec{\op{S}}$, while the back action from $\vec{\op{S}}$ to $\vec{\op{J}}$ is rather small, which we expect as well for our dissipative case. Poincar{\'e} surfaces of section for the motion of the small spin suggest that there exist regions in ($B,\lambda$) parameter space with a mixed phase space for the small spin, i.e., regular or chaotic motion for different initial conditions, even while the LS's motion remains regular.

The analogies between classical and quantum chaotic systems of interacting spin clusters have been investigated in terms of quantum webs of simultaneous eigenstates that can be used to illustrate regular as well as chaotic dynamics~\cite{Srivastava1990,Robb1998}.

\subsection{\label{sec:infinite_bias}Dynamics in the Infinite Bias Regime}

Throughout this analysis we consider an anisotropic coupling between the spins, namely, $\lambda_y = 0$ and $\lambda_x = \lambda_z = \lambda$ (compare Refs.~\onlinecite{Lopez-Monis2012,Metelmann2012}). The qualitative system dynamics for double and single QD systems with isotropic coupling has been investigated in Ref.~\onlinecite{Mosshammer2012}. For the sake of clarity we choose the QD level to be $\varepsilon = 0$. Furthermore, we choose the leads to be at equal temperatures $\beta$. The chemical potentials allow for tunneling from left to right only ($\mu_{L,R} \to \pm \infty$).

In the following we discuss the difficult nonlinear dynamics occurring in our setup. In particular, we are interested in the average spin $\sigma$-electron current through the barrier $l$, i.e.,
\begin{align}
\label{eq:current_through_lbarrier}
 \expval{I_{l\sigma}}(t) &= e \Gamma_{l\sigma} \Big[\e^{\delta_{l\sigma}} \overline{f_{l\sigma}} \expval{\op{n}_\sigma} \\
&\quad - \e^{\delta_{lE}} f_{l\sigma} \left(1 - \expval{\op{n}_\sigma} - \expval{\op{n}_{\nsigma}} \right) \Big] \,, \nonumber
\end{align}
where, by convention, net flux off the SQD is positive. Further, we are interested in the backaction on the LS and what the influence of the applied feedback mechanisms is.
As stated above, we consider quantum expectation values of the electronic spin's components $\expval{\op{S}_i}$, while the LS $\vec{\op{J}}$ is treated as a classical object, which is justified and motivated as we remain in a limit where $s \ll j$.
\subsubsection{Analytic fixed points}
To describe the domain of regular motion it is of particular interest to investigate whether there are fixed points. We denote the fixed points introducing the notation
$ \calp = \left( \expval{\op{S}_x^*},\expval{\op{S}_y^*},\expval{\op{S}_z^*},\expval{\op{J}_x^*},\expval{\op{J}_y^*},\expval{\op{J}_z^*} \right)$. 
For a general choice of parameters it might become difficult to calculate the fixed points straightforwardly, since the introduction of the Fermi function renders the equations transcendental. For special infinite-bias setups, nonetheless, it is possible to calculate them directly, by setting Eqs.~\eqref{eq:eom_sqd} to zero and solving them.

However, due to the nature of the exchange interaction one can easily see that there exist trivial fixed points when the LS is aligned (anti-)parallel with the magnetic field since the spins decouple then. 
If there is no coupling to the leads, the electronic spin is conserved and will be aligned with the magnetic field. However, contrasting the spin-conserving setup we have to take into account what happens to the electronic spin, once the exchange interaction is ineffective. This, of course, depends on the dissipative setup, i.e., $\Gamma_{l\sigma}$. The most trivial dissipative setup consists of setting all tunneling rates to $\Gamma$ and applying an infinite-bias voltage, so that the electronic spin will simply decay, all further electrons tunnel through the device without being flipped, and the transport channels are decoupled. 

If the setup is changed in a way that tunneling between the right contact and the SQD is only possible for $\spup$ electrons, in terms of the tunneling rates $\Gamma_{R\spdo} = 0, \Gamma_{L\spup} = \Gamma_{R\spup} = \Gamma_{L\spdo} = \Gamma$, $\spdo$ electrons will be trapped in the system and can only leave the SQD after being flipped. Note, that in this setup (labeled ``IB''), which we consider throughout this section, the two feedback schemes introduced above are identical, which can be verified by applying the scheme to Eqs.~\eqref{eq:eom_sqd}, and thus the feedback strength enters the equation by the dimensionless parameter $\delta$.

The respective fixed points are, accordingly,
\begin{align}
 \calp_{\mbs{IB}}^{\pm} &= \left(0,0,-\frac{1}{2},0,0, \pm j \right)\,.
\end{align}
These fixed points are independent of the parameters $B,\lambda$ and, therefore, exist in the whole parameter regime. The fixed points can be investigated further using linear stability analysis. Such analysis requires the knowledge of the partial derivatives of spin components with respect to other components, which are, in general, not accessible.

There is a second set of fixed points,
\begin{align}
\calp_{\mbs{IB}}^{-y,\pm} &= \left(0,- \calb_3 ,-\frac{B}{\lambda}, - \frac{\Gamma}{2 B} \calb_3, \pm \calb_2,-\frac{B}{\lambda} \right)\,, \nnn
\calp_{\mbs{IB}}^{+y,\pm} &= \left(0,+ \calb_3 ,-\frac{B}{\lambda}, + \frac{\Gamma}{2 B} \calb_3, \pm \calb_2,-\frac{B}{\lambda} \right)\,,
\end{align}
with
\begin{align}
\calb_2 &= \sqrt{j^2-\left(\frac{B}{\lambda} \right)^2 - \frac{\e^{\delta}}{5} \left(\frac{\Gamma}{\lambda}\right)^2 \left(\frac{\lambda}{2B} -1  \right)} \,, \nnn
\calb_3 &= \e^{\delta/2}\frac{2}{\sqrt{5}} \sqrt{\frac{B}{\lambda} \left(\frac{1}{2} - \frac{B}{\lambda} \right)}\,.
\end{align}
This fixed point can be understood in terms of the anisotropic exchange coupling, since the $x$- and $z$ components of one and the same spin decouple, having either the trivial consequence of complete spin polarization ($\calp_{\mbs{IB}}^{\pm}$) or the $z$ components being determined by the ratio $-B/\lambda$ and vanishing $x$ components. 
%

These results are very similar to the findings in Refs.~\onlinecite{Lopez-Monis2012,Metelmann2012}, only differing due to different choices of the transport rates and the introduction of the feedback parameter. Note that we explicitly restricted the number of allowed electrons in the SQD at a time to one, which is not the case for Ref.~\onlinecite{Lopez-Monis2012}. 

The quantities $\calb_2$, $\calb_3$ can assume finite imaginary values. The fixed points $\calp_{\mbs{IB}}^{\pm y,\pm}$, therefore, only have a physical meaning in the region of the parameter space where they remain real valued. The parameter space can thus be separated into different regions whose boundaries are obtained by solving $\calb_2 = 0$ and $\calb_3 = 0$. We obtain the critical values
\begin{align}
 \calb_2 &= 0 \rightarrow \Gamma_c = \e^{-\delta/2} \sqrt{10 \frac{(B^3-B j^2 \lambda^2)}{2 B - \lambda}} \,, \nnn
 \calb_3 &= 0 \rightarrow B_c = \frac{\lambda}{2} \,.
\end{align}

\begin{figure}[t!]
\label{fig:param_regions_infinite_bias}
 \includegraphics[width=0.48\textwidth,keepaspectratio=true]{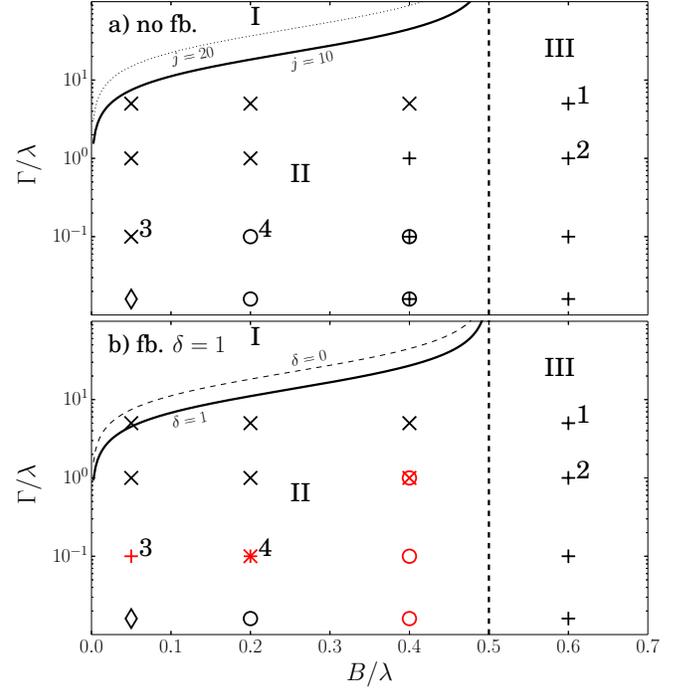}
 \caption{(Color online) Parameter regions I,II and III in the infinite-bias (IB) setup. Solid lines show the region-I-II-border [as a reference the borders for larger $j$ (dotted) in a) or the nonfeedback case (dashed) in b) are shown]. The thick dashed lines show the transition from region II to region III, which is determined by the critical magnetic field $B_c=\lambda/2$.
 In region I the system always runs into either of the fixed points $\calp_{\mbs{IB}}^\pm$ performing damped oscillations. If the magnetic field is stronger than $\lambda/2$ (region III) the system performs self-sustained oscillations, either according to the effective model \eqref{eq:effective_model} or around the polar fixed points $\calp_{\mbs{IB}}^\pm$. For region II, we plot exemplarily which dynamics the system exhibits with respect to parameters $B$ and $\Gamma$. We find not only damped oscillations towards the system's fixed points (marked by ``$\times$'') and ongoing oscillations (``$+$'') but also quasiperiodic (``$\circ$'') or chaotic oscillations (``$\diamond$''). 
 Plot a) shows the typical behavior for the nonfeedback case. In plot b) we show how and in which parameter regions the system's behavior is changed qualitatively if feedback is applied ($\delta/\lambda = 1$). The red labels show the regions in parameter space, where the dynamics are changed significantly. Labels 1--4 show the parameters for which we plotted solutions in Figs.~\ref{fig:transitions_regIII_IB2} to \ref{fig:transitions_regII_B0.2}.}
\end{figure}

A projection of the three-dimensional parameter space ($\lambda,B,\Gamma$) for different feedback parameters $\delta$ with fixed $\lambda$ on the $\Gamma-B$ plane is, therefore, divided into three regions as plotted in Fig.~\ref{fig:param_regions_infinite_bias}. $\calp_{\mbs{IB}}^{\pm}$ are physical fixed points for all possible parameters.

The labels 1 to 4 in Fig.~\ref{fig:param_regions_infinite_bias} mark the parameters for which we solve the numerics and plotted the results in Figs.~\ref{fig:transitions_regIII_IB2} to \ref{fig:transitions_regII_B0.2}. The specific dynamics are discussed in the following.

\subsubsection{Region I: damped oscillations}
In region I with weak magnetic fields $B < B_c$ and very fast tunneling processes, i.e., rates $\Gamma > \Gamma_c$, the time scale of electronic transport is much faster than that of the exchange interaction; therefore, the rate equations results only show damped oscillations in the current and evolution towards $\calp_{\mbs{IB}}^\pm$. The resulting stationary current will be obviously zero, since the spin-spin interaction is ineffective once either of the fixed points is reached and a $\spdo$ electron is trapped in the QD.
The origin of the often complicated transient oscillations is the interplay between the modulation of the electronic current by the LS and the electronic feedback on the LS. Different initial conditions decide which of the two fixed points is reached. Turning on the demon feedback does not change the behavior qualitatively, but only the shape of the transient oscillations; their frequency and the damping is increased.

\subsubsection{Region III: self-sustained oscillations, parametric resonance and an effective model}
If the magnetic field is increased beyond $B_c$ (parameter region III), however, the numerical results exhibit solely self-sustained oscillations in the electronic current and both electronic and LS.

The system oscillates on orbits around the stable fixed points $\calp_{\mbs{IB}}^{\pm}$, with the LS performing periodic oscillations close to full polarization $\pm j$, driving the electronic spin to a periodic orbit close to $\expval{\op{S}_z^*} = - 1/2$. 

A second type of self-sustained oscillation can be described by an effective model, where $\expval{\op{J}_z} = -B/\lambda$, and $\expval{\op{S}_x} = 0$ remain constant and the oscillation of $\expval{\op{J}_{x,y}}$ is driving the periodic oscillation of the small spin, which reduces the system to
\begin{align}
\label{eq:effective_model}
\frac{d}{dt} \expval{\op{S}_x} &= -\frac{\Gamma}{2} \expval{\op{S}_x} \nnn
\frac{d}{dt} \expval{\op{S}_y} &= - \frac{1}{2} B_x(t)\left(\expval{\op{n}_\spup} - \expval{\op{n}_\spdo} \right) -\frac{\Gamma}{2} \expval{\op{S}_x} \\
\frac{d}{dt} \expval{\op{n}_\spup} &= B_x(t) \expval{\op{S}_y} + \e^{\delta} \Gamma -\e^{\delta} \Gamma \left(2\expval{\op{n}_\spup} + \expval{\op{n}_\spdo} \right)\nnn
\frac{d}{dt} \expval{\op{n}_\spdo} &= - B_x(t) \expval{\op{S}_y} + \e^{\delta} \Gamma -\e^{\delta} \Gamma \left(\expval{\op{n}_\spup} + \expval{\op{n}_\spdo} \right)\,, \nn
\end{align}
with the oscillating effective magnetic $B_x(t) = \frac{1}{\sqrt{2}} \sqrt{j^2-\frac{B^2}{\lambda^2}} \lambda \left(\cos\left(B_{\mbs{eff}} t\right) - \sin\left(B_{\mbs{eff}} t\right)\right)$. Thus, the oscillation of the electronic plane is taking place in the $y-z$ plane, while the LS rotates in the $x-y$ plane. The derivation of this model in Appendix~\ref{sec:derivation_effective_model} makes use of the rotational invariance of the LS. The effective precession frequency, $B_{\mbs{eff}}$ is decreasing with increasing $\Gamma$, as well as with increasing $\delta$. Due to the term $B + \lambda \expval{\op{S}_z}$ and the fact that $\expval{\op{S}_z} < 0$ is, on average, the effective precession frequency $B_{\mbs{eff}}$ is smaller than $B$ for all cases we examined. We find $B_{\mbs{eff}} \approx B$ for very small $\Gamma$.
The same consideration gives rise to the discovery that for the orbits around the fixed points $\calp_{\mbs{IB}}^{\pm}$ the effective precession frequency is much smaller than $B$. 

A numerical Fourier analysis reveals the frequencies in the spins' motions for the periodic oscillations: We identify the frequencies $\omega_s$, which appears dominantly in the small spins motion, and $\omega_j$, which drives the LS. Due to the exchange interaction the electronic spin also assumes $\omega_j$ as an enveloping frequency. 
Since we consider particularly setups, where the magnitudes of the spins differ largely ($j \gg s$), we find $\omega_s \gg \omega_j$.  For the $z$ components we find a frequency doubling $\omega_z = 2 \omega_j$ as well as additional differential frequencies $\omega_\pm = \omega_s \pm \omega_j$. Those frequencies can be understood as the $\expval{\op{S}_z}$ component does not couple to the magnetic field but is a product of $\expval{\op{J}_x}$ and $\expval{\op{S}_y}$.
Ideally, those components develop as a superposition of sinusoidal functions of frequencies $\omega_s$ and $\omega_j$, namely, $\expval{\op{J}_x^{\mbs{eff}}(t)} = b_s \sin \left(\omega_s t + \theta_s \right) + b_j \sin \left(\omega_j t + \theta_j \right)$ and $\expval{\op{S}_y^{\mbs{eff}}(t)} = a_s \sin \left(\omega_s t + \phi_s \right) + a_j \sin \left(\omega_j t + \phi_j \right)$, respectively. The frequency $\omega_j$ describes an enveloping oscillation of the faster $\omega_s$ oscillations, while, typically the backaction from $\vec{\op{S}}$ to $\vec{\op{J}}$ is very small ($b_s \ll b_j $). Assuming, further, in-phase oscillations with respect to $\omega_j$, i.e., $\phi_j = \theta_j = 0$, formal integration ($\lambda = 1$) leads to
\begin{align}
 \expval{\op{S}_z^{\mbs{eff}}(t)} &=  \int_0^t dt' \expval{\op{S}_y^{\mbs{eff}}(t')} \expval{\op{J}_x^{\mbs{eff}}(t')} \nnn
 & = - \frac{a_j b_j \sin \left(2 \omega_j t \right)}{4 \omega_j} + \frac{a_s b_j \sin{(\phi_s - \omega_- t)}}{2 \omega_-}  \nnn
 & + \frac{a_s b_j \sin{(\phi_s + \omega_+ t)}}{2 \omega_+} + C\,.
\end{align}
The exact frequencies and amplitudes depend on the system parameters. The larger the dissipation gets, the more the electronic spin assumes the LS's frequency, which is due to the faster decay of SQD states. The initial conditions of the LS also play an important r\^ole since the exchange interaction is enhanced the more $\vec{\op{J}}$ is perpendicular to the magnetic field.

\begin{figure}[t!]
 \label{fig:transitions_regIII_IB2}
 \includegraphics[width=0.48\textwidth,keepaspectratio=true]{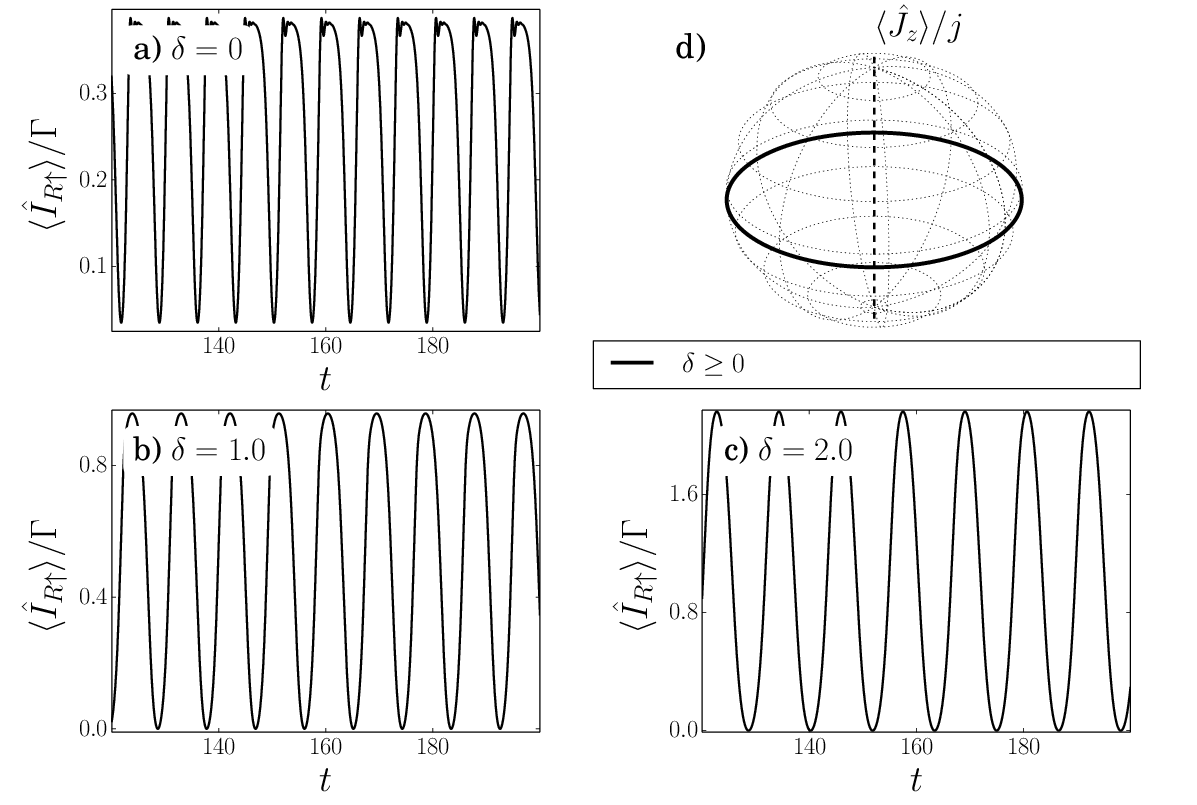}
 \caption{a)--c) Electronic currents and d) the dynamics of $\vec{\op{J}}$ for parameters from region III with respect to different feedback strengths $\delta$. The current becomes more sinusoidal, while the frequency is decreased; the rotation of $\vec{\op{J}}$ remains in the $x-y$-plane.  Parameters: $B/\lambda = 0.6, \Gamma/\lambda = 5, j/\lambda = 10$.} 
\end{figure}

\begin{figure}[t!]
 \label{fig:transitions_regIII_IB1}
 \includegraphics[width=0.48\textwidth,keepaspectratio=true]{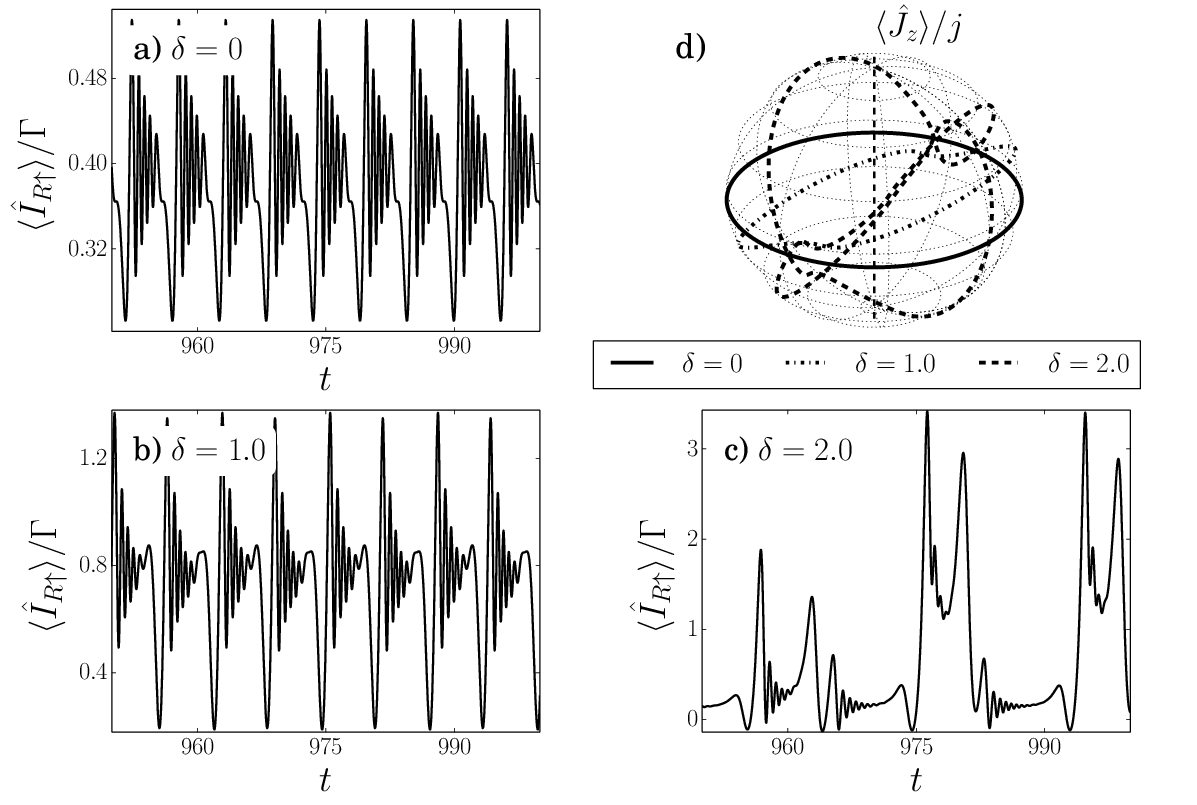}
 \caption{a)--c) Electronic currents and d) the dynamics of $\vec{\op{J}}$ for parameters from region III with respect to different feedback strengths $\delta$. The current oscillations become slower and turn to quasiperiodic oscillations. The oscillation of $\vec{\op{J}}$ is not planar anymore and the limit cycles become more complicated. Parameters: $B/\lambda = 0.6, \Gamma/\lambda = 1, j/\lambda = 10$.}
\end{figure}

Note also that for the same reasons a frequency doubling for the oscillation of the $z$ components of the spins and the electronic currents occur for every periodic oscillation, damped and undamped. Before reaching stationary cycles the system often undergoes transient oscillations.

As long as the feedback mechanism is switched off this holds for the whole parameter region. Once the feedback is switched on, for instance, with a feedback constant $\delta$ in the order of magnitude of $\lambda$, region III exhibits subregions with respect to the magnitude of $\Gamma$.

For $\Gamma \gg \lambda$ the LS oscillates in the $\expval{\op{J}_z^*} = -B/\lambda$ plane, independent of $\delta$. The corresponding currents are regularized to become almost sinusoidal; cf. Fig.~\ref{fig:transitions_regIII_IB2}. 

With much slower electronic transport $\Gamma \ll \lambda$ the oscillation plane of $\vec{\op{J}}$ is tilted by an angle $\phi$ around the $x$ axis and we find even more complicated orbits for the oscillation of $\vec{\op{J}}$ that is not restricted to a plane anymore if $\Gamma$ is close to $\lambda$. Since $\delta$ enters exponentially into the equations the tunneling setup is, thus, detuned drastically if the feedback parameter $\delta$ is increased. For some parameters the dynamics change towards quasiperiodic oscillations or even chaos.
In Fig.~ \ref{fig:transitions_regIII_IB1} we have plotted the currents and the LS's cycles for different $\delta$. We see that the regular motion becomes quasiperiodic for increasing feedback strength. For intermediate $\delta$ the $\vec{\op{J}}$ rotation stays planar, but the rotation plane is tilted.
This kind of limit cycle oscillation where the plane of oscillation through $\expval{\op{J}_z} = -B/\lambda$ is tilted by an angle $\phi$ around the $x$ axis may be described by an effective model similarly to \eqref{eq:effective_model}, if one found the angle $\phi$ and adjusted the transformation in Appendix~\ref{sec:derivation_effective_model} accordingly.

\begin{figure}[t!]
 \label{fig:compare_eff_model}
 \includegraphics[width=0.49\textwidth,keepaspectratio=true]{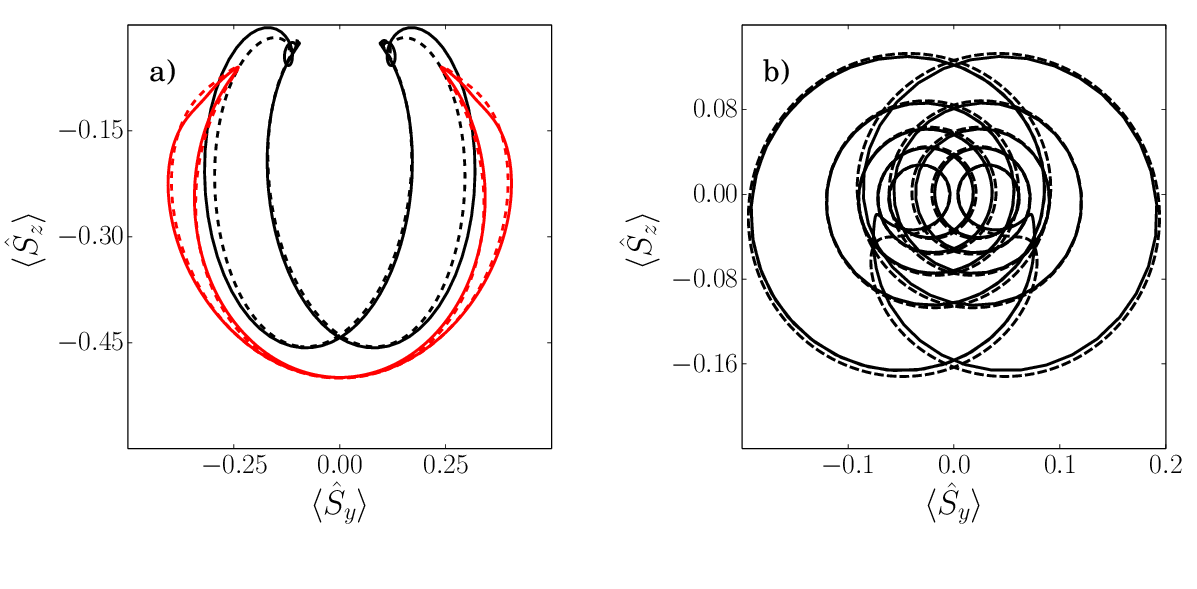}
 \caption{(Color online:)Comparison between solutions to the full set of Eqs.~\eqref{eq:eom_sqd} and \eqref{eq:eom_largespin_expanded} (dashed lines) and solutions to the effective model \eqref{eq:effective_model} (solid lines) for different setups. Parameters a) $B/\lambda = 0.6$, $\Gamma/\lambda=5$, $j/\lambda = 10$, $\delta/\lambda = 1$ (red) and $\delta = 0$ (black). Parameters b) $B/\lambda = 0.6$, $\Gamma/\lambda=1$, $j/\lambda = 10$, $\delta=0$.}
\end{figure}

In Fig.~\ref{fig:compare_eff_model} we compare the solutions for our effective model to the full solutions for different $\Gamma$ and, if possible, for different $\delta$. 

One further remark should be given on the periodic orbits. The oscillation around the polarization fixed point $\calp_{\mbs{IB}}^{\pm}$ is undamped and the magnitude of the electronic spin is nearly constant. In this case the two interacting spins can be considered as a system of two interacting undamped Bloch equations. Those could be recast in the form of coupled complex Riccati equations~\cite{Hasegawa1986,Kobayashi2004a}. Unfortunately, to the best of our knowledge these Riccati equations can only be solved for special setups, e.g., isotropic coupling.

\subsubsection{Region II: mixed dynamics}
We observe the full variety of the possible dynamics, in regime II. All fixed points are real valued, but $\calp_{\mbs{IB}}^{\pm}$ are unstable while  $\calp_{\mbs{IB}}^{\pm y,\pm}$ are stable. We find many parameter settings and initial conditions where the system will oscillate strongly damped to run into $\calp_{\mbs{IB}}^{\pm y,\pm}$. Once one of the fixed points $\calp_{\mbs{IB}}^{\pm y,\pm}$ is reached, a stationary current is established that follows straightforwardly from \eqref{eq:eom_sqd}. We find the stationary state of the total SQD occupancy $\op{N}=\sum_\sigma \op{n}_\sigma$ by solving 
\begin{align}
 0 &= -\e^{\delta} \Gamma \left(5 \expval{\op{n}^*_\spup} - 4 \expval{\op{S}_z^*} - 2 \right)\,,
\end{align}
which provides the stationary current of $\spup$ electrons leaving the SQD ($\expval{\op{S}_z^*}\to - B/\lambda$)
\begin{align}
\label{eq:IB_fpy_current}
 \frac{\expval{I_{R\spup}}}{e \Gamma} &= \frac{2}{5}\e^{\delta} \left( 1-2\frac{B}{\lambda} \right)\,.
\end{align}
On the other hand, the current will vanish when the system reaches the full polarization fixed point $\calp_{\mbs{IB}}^{\pm}$, as the SQD is found in a Coulomb blockade then.

Additionally, we find quasiperiodic or even chaotic oscillations and two kinds of self-sustained oscillations as we have found for region III.

If the feedback mechanism is switched on and the system is started from the same initial conditions we often find the system to respond in a way that an oscillation vanishes and the system runs into fixed points $\calp_{\mbs{IB}}^{\pm y,\pm}$. Periodic oscillations for magnetic fields close to $B_c$ are often changed to run into fixed points or detuned into quasiperiodic oscillations.

\begin{figure}[t!]
 \label{fig:transitions_regII_B0.05}
 \includegraphics[width=0.48\textwidth,keepaspectratio=true]{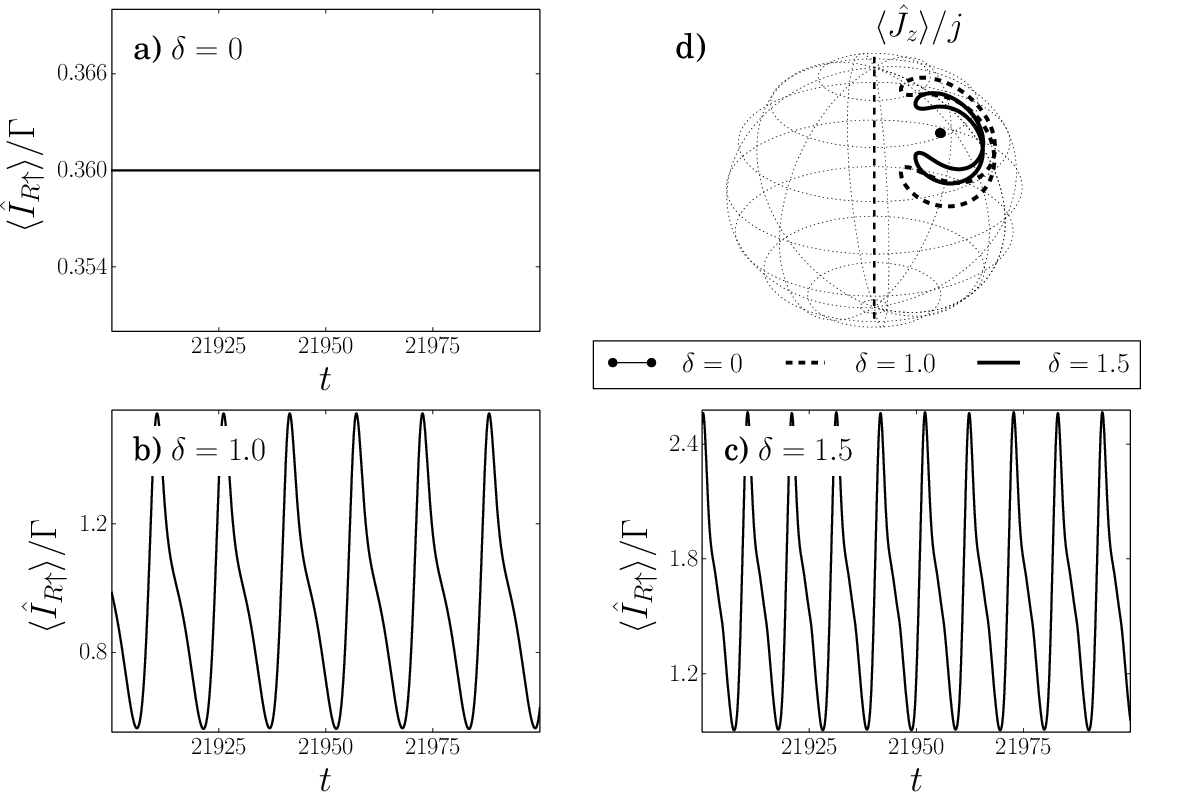}
 \caption{a)--c) Electronic currents and d) the dynamics of $\vec{\op{J}}$ for parameters from region II with respect to different feedback strengths $\delta$. The evolution towards the fixed point $\calp^y$ [with a corresponding current \eqref{eq:IB_fpy_current}] is changed into self-sustained oscillations via feedback. Parameters: $B/\lambda = 0.05, \Gamma/\lambda = 0.1, j/\lambda = 10$.}
\end{figure}

\begin{figure}[t!]
 \label{fig:transitions_regII_B0.2}
 \includegraphics[width=0.48\textwidth,keepaspectratio=true]{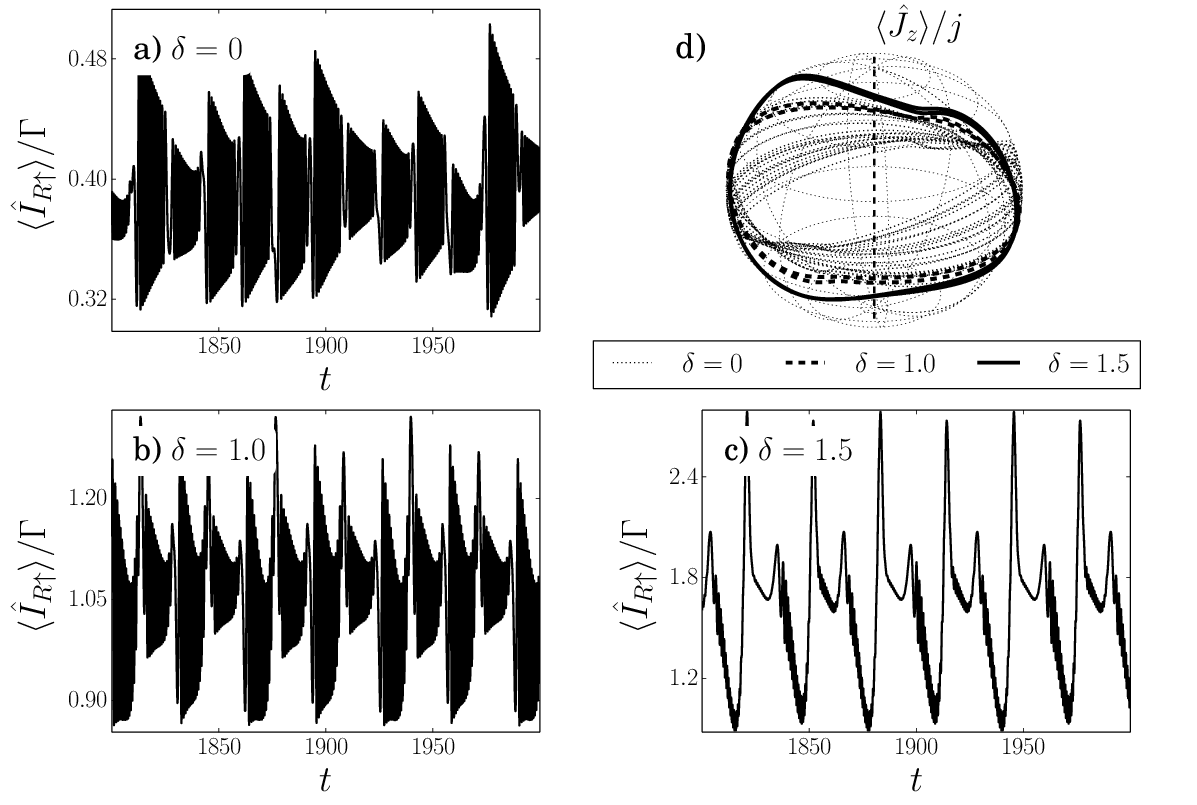}
 \caption{a)--c) Electronic currents and d) the dynamics of $\vec{\op{J}}$ for parameters from region II with respect to different feedback strengths $\delta$. Irregular oscillations are regularized for both electronic and LS. Parameters: $B/\lambda = 0.2, \Gamma/\lambda = 0.1, j/\lambda = 10$.}
\end{figure}

Nonetheless, the numerical results show that there are parts of the parameter space where feedback turns quasiperiodic oscillations into regular self-sustained, though not planar, oscillations; compare Fig.~\ref{fig:transitions_regII_B0.2} and labels 3 and 4 in Fig.~\ref{fig:param_regions_infinite_bias}. It is possible as well to prevent the system from running into its fixed points and induce periodic oscillations instead; compare Fig.~\ref{fig:transitions_regII_B0.05}.


Chaotic behavior in region II is dominant in parameter regimes where the exchange interaction is much stronger than the tunnel coupling, i.e., $\Gamma \ll \lambda$ and small magnetic field $B \ll \lambda$. As a consequence the feedback effect is rather small in these regimes, at least with feedback parameters $\delta$ in the order of magnitude of $\lambda$. 


\subsection{\label{sec:finite_bias}Finite bias regime: fixed points and parameter regions}
In the finite-bias (FB) regime we choose the tunneling rates asymmetrically as or the infinite-bias setup, because of the richness of the occurring dynamics. In the following we assume a symmetric detuning of the chemical potentials with respect to the SQD level ($\varepsilon=0$), i.e. $\mu_L = -\mu_R = V/2$. In the limit of vanishing temperatures only electrons with energies between the leads' chemical potentials can participate in transport. This so-called transport window is smeared out due to thermal melting of the Fermi edges if the temperatures are finite. We, however, consider equal finite temperatures $\beta_l = \beta$ and choose the bias voltage in a manner that for given parameters $B,\lambda,j$ the energies $\varepsilon_\sigma(t)$ lie between the chemical potentials for all times. The maximum detuning between the $\spup$ and $\spdo$ levels is $\varepsilon_z^{\mbs{max}} = B + \lambda_z j$, and accordingly we choose $ \beta V > \beta \varepsilon_z^{\mbs{max}}$.

We have to remark that bidirectional tunneling for this parameters is only possible for finite temperatures, obviously.

\subsubsection{Fixed points}
\begin{figure}[t!]
\label{fig:statcurrent_polfixpoint}
\includegraphics[width=0.48\textwidth,keepaspectratio=true]{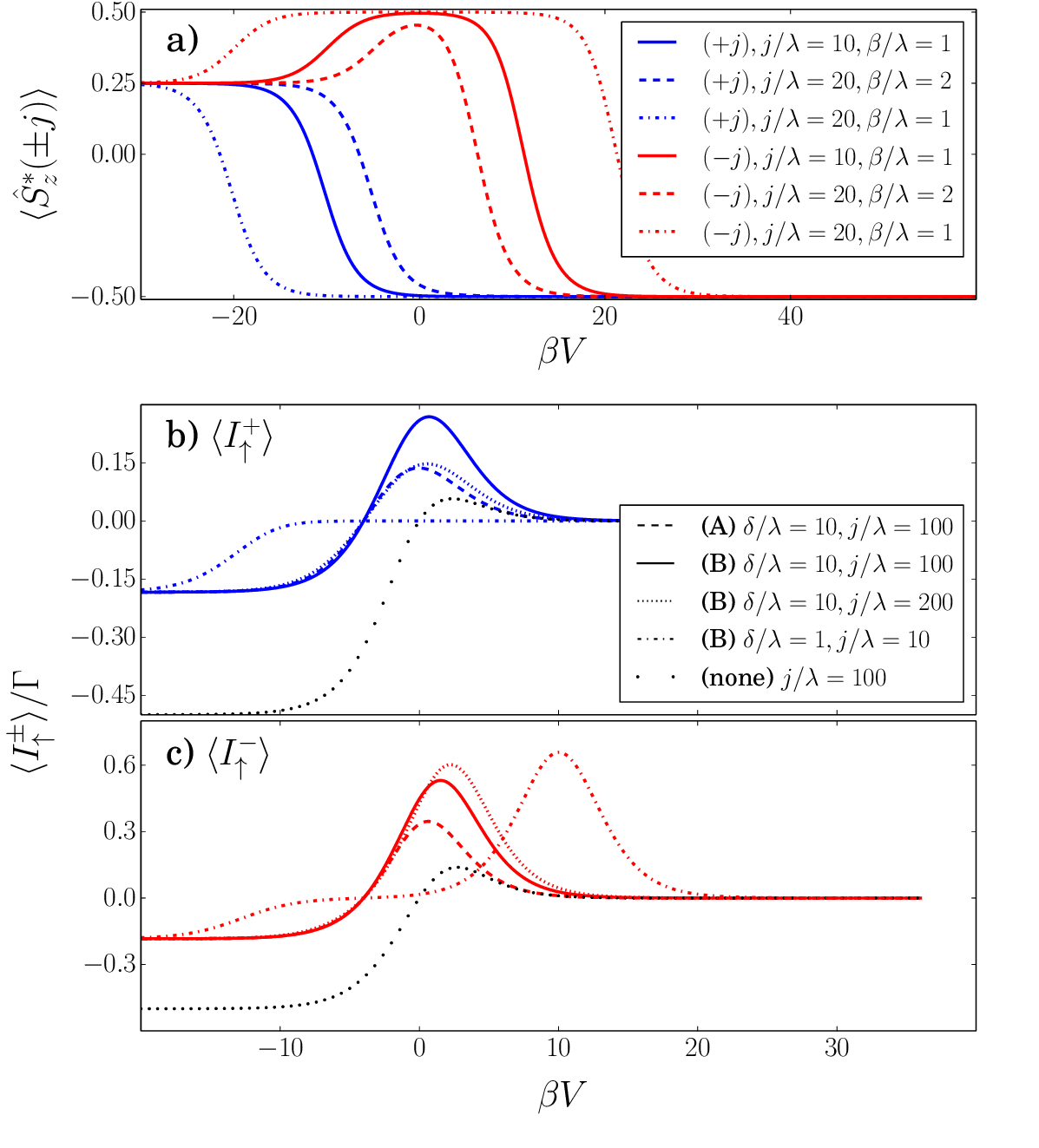}
\caption{(Color online) a) The values of the $S_z$ component of the global fixed  point $\calp_{\mbs{FB}}^{\pm}$ vs the bias voltage.
The spin-dependent stationary currents after the system reaches the fixed point $\calp_{\mbs{FB}}^+$ (b) or $\calp_{\mbs{FB}}^-$ (c) are shown for feedback schemes A and B and the nonfeedback case.}
\end{figure}

As in the previous sections, we find fixed points, where the dynamics of the spins are decoupled and $\vec{\op{J}}$ is fully polarized,
\begin{align}
 \calp_{\mbs{FB}}^{\pm} &= \left(0,0,+\calb_{4\pm}^{\mbs{A,B}}, 0, 0, \pm j \right)\,
\end{align}
The values $\calb_{4\pm}^{\mbs{A,B}}$ are dependent on the respective feedback scheme. They are denoted in Appendix \ref{sec:app_finitebias_fixedpoints}. The alignment of the electronic spin is not only asymmetric with respect to the detuning of the leads' chemical potentials, but also with respect to the LS polarization, i.e. for increasing $\lambda$ and a large $j$ we find trapped electrons with its spin antiparallel to the polarization of the LS: $\expval{\op{S}_z^*}(\pm j,\lambda \gg 0 , V = 0) \to \mp 1/2$. In Fig.~\ref{fig:statcurrent_polfixpoint} we plot $\expval{\op{S}_z^*}$ and the stationary states that are established once the spins decouple; compare Appendix \ref{sec:app_finitebias_fixedpoints}.




As in the infinite-bias setup there is a second set of fixed points, with $\expval{\op{S}_z^*} = \expval{\op{J}_z^*} = -B/\lambda$, namely,

\begin{align}
 \calp_{\mbs{FB}}^{\pm y, \pm} &= \Bigg(0,\pm \sqrt{\frac{2B}{\lambda}} \frac{1+\e^{\frac{\beta V}{2}}}{2+\e^{\frac{\beta V}{2}}} \calb_6^{\mbs{A/B}},-\frac{B}{\lambda}\,, \nnn
 & \quad \quad \pm \frac{\Gamma}{\sqrt{2 B \lambda}} \calb_6^{\mbs{A/B}} ,\pm \calb_7^{\mbs{A/B}} , -\frac{B}{\lambda} \Bigg)\,,
\end{align}
with the parameters
\begin{align}
\calb_6^{\mbs{A}} &= \sqrt{-\frac{\left(4 B+\e^{\delta + \frac{\beta V}{2}}(2B-\lambda) + \lambda \right)}{\lambda\left(3 + \e^\delta + 5 \e^{2\delta + \frac{\beta V}{2}} \right)} C} \,, \nnn
\calb_6^{\mbs{B}} &= \sqrt{-\frac{\e^\delta \left(4 B+\e^{ \frac{\beta V}{2}}(2B-\lambda) + \lambda \right)}{\lambda\left(3 + \e^{2\delta}(1 + 5 \e^{\frac{\beta V}{2}}) \right)} C} \,, \nnn
\calb_7^{\mbs{A/B}} &= \sqrt{j^2 -\left(\frac{B}{\lambda}\right)^2 - \frac{\Gamma^2}{2 B \lambda} (\calb_6^{\mbs{A/B}})^2}\,, \nnn
C &=  (2+\e^{\frac{\beta V}{2}})\frac{1+\e^{2 \delta + \frac{\beta V}{2}}}{(1+\e^{\frac{\beta V}{2}})^2} \,.
\end{align}

We can, once again, identify a region, where the fixed points $\calp_{\mbs{FB}}^{\pm y, \pm}$ assume finite real values for all components (region II) and regions (I,III) where they have no physical meaning. We solve the equations $\calb_6 = 0$ and $\calb_7 =0$ and obtain the critical values
\begin{align}
B_c^{\mbs{A}} &= \frac{\lambda}{2}\left(1-\frac{3}{2 + \e^{\delta+\frac{\beta V}{2}}} \right)\,, \nnn
B_c^{\mbs{B}} &= \frac{\lambda}{2}\left(1-\frac{3}{2 + \e^{\frac{\beta V}{2}}} \right)\,, \\
\Gamma_c^{\mbs{A}} &= \sqrt{ \frac{2 B \left(5 \e^{2 \delta+\frac{\beta V}{2}}+\e^{\delta}+3\right) (B-\lambda j) (B+\lambda j)}
{\left((2 B-\lambda)\e^{\delta+\frac{V}{2}}+4 B+\lambda\right)} \frac{1}{C}}\,, \nnn
\Gamma_c^{\mbs{B}} &= \sqrt{ \frac{2 B \left(5 \e^{2 \delta+\frac{\beta V}{2}}+\e^{2\delta}+3\right) (B-\lambda j) (B+\lambda j)}
{\e^{\delta} \left((2 B-\lambda)\e^{\frac{V}{2}}+4 B+\lambda\right)} \frac{1}{C} }\,. \nonumber
\end{align}
Interestingly, the border between regimes II and III for the feedback scheme B is the same as for the nonfeedback case. 
Thus, switching between the two considered feedback schemes might go along with transitions between the parameter regions. 

The dynamics for the three parameter regions are quite similar to that in the infinite-bias regime. Though, there are some special features.


The dominant behaviors in region I are damped oscillations towards the polarization of spins $\calp_{\mbs{FB}}^{\pm}$. There exist additional fixed points where $\expval{\op{J}_y^*} \to 0$ for intermediate bias voltages. It is not possible to calculate them analytically, since $\expval{\op{J}_z}$ couples to $\expval{\op{S}_x}$ and $\expval{\op{S}_y}$ via hyperbolic functions. 

In region II the system can -- depending on the tunneling rate $\Gamma$ and the initial conditions -- run into one of the fixed points $\calp_{\mbs{FB}}^{\pm y, \pm}$. For smaller $\Gamma$ the spins may as well perform periodic oscillations or even chaotic ones, similar to the behavior for the infinite-bias case.

We find dynamics that are very similar to the infinite-bias case in region III. Nevertheless, additional oscillations occur, where $\vec{\op{J}}$ is only partially polarized antiparallel to the external magnetic field and its magnitude is increased with increased bias voltage. This partial polarization is accompanied by the decay of the the electronic states and coherences and, thus, constant average currents. It takes place for intermediate bias voltage of $\beta V/\lambda \approx 5$.
If the bias voltage is increased and given suitable initial conditions the system ends up completely polarized and electrons get trapped while the energy level for electrons of the opposite spin is shifted outside the transport window.
The partial polarization is $P_\kappa = \vert \expval{\op{J}_z^{*\kappa}} \vert$ with $\kappa$ denoting the feedback scheme A,B, or 0 (no feedback applied). It is of magnitude $P_\kappa/\beta V \approx 1$ and is unique for all tuples ($B,\lambda,V,\delta$) and the respective feedback scheme and we find $P_{\mbs{A}} > P_{\mbs{B}} > P_{0}$.

\subsection{\label{sec:discussion_trajectory}Typical feedback trajectories, spin filter setups}
\begin{figure}[t!]
\label{fig:sv_currents_trajecs_both}
\includegraphics[width=0.48\textwidth,keepaspectratio=true]{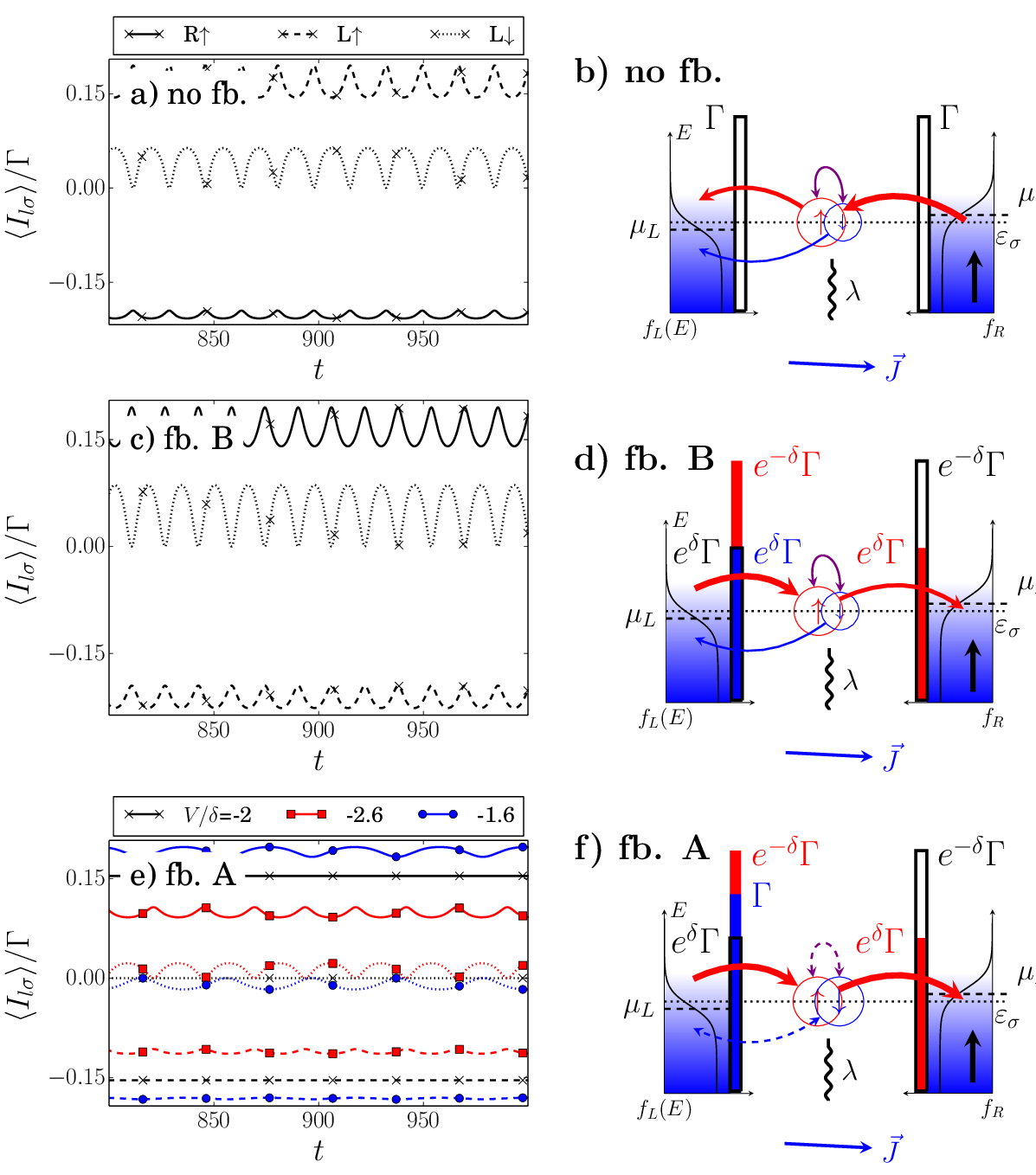}
\caption{(Color online) Average spin-$\sigma$ currents through barrier $l$ for small negative biases and different feedback schemes. Without feedback (a) the currents flow according to the applied bias voltage from the right lead to the left lead. For feedback scheme B c) an oscillating $\spup$ current against the bias is building up, and $\spdo$ electrons are ejected with the bias, the device is acting as a spin filter. With the given parameters the two transport channels decouple for feedback scheme A e). A little detuning of the bias voltage leads to the oscillating $\spup$ currents against the bias. $\spdo$ electrons are either ejected to the left reservoir(red squares) or trapped (blue circles). Images b),d) and f) show the SQD with the respective tunneling rates if the system state is empty (black), $\spup$ (red) or $\spdo$ (blue). Thickness and direction of the arrows show which net currents are reached. Level transitions (spin flips) are shown in violet. In f) the transport channel decouple, no spin flips 
occur, and the $\spdo$-current vanishes (dashed lines).  Parameters: $B/\lambda = 0.1$, $\Gamma/\lambda=10$,$\delta/\lambda=10$ and $|V|/\lambda = 20$.}
\end{figure}

In the following we want to sketch typical trajectories found for a spin filter setup that makes use of the demon feedback. As stated above, the setup without interaction with the external spin generates currents even against small bias voltages (compare Sec.~ \ref{sec:closed_system_dynamics}). To demonstrate the spin filtering effect, while the interaction with the LS is effective we suggest a small negative bias voltage with $|V| > B + \lambda j$. For small ratios $\lambda/j$ the electronic backaction on $\vec{\op{J}}$ is very small and for suitable initial conditions the LS stabilizes at $\expval{\op{J}_z} = -B/\lambda$ and precesses freely. In this special case the energies $\varepsilon_\sigma$ coincide and become constant in time. Due to the anisotropic exchange interaction the occupations of the $\spup$  and $\spdo$ levels oscillate with the doubled precession frequency, and we find regularly oscillating currents accordingly.

If no feedback takes place, the currents are flowing with the bias, i.e., from the right to the left. Due to the exchange interactions the averaged $\spup$ influx through the right lead splits up to generate oscillating $\spup$ and $\spdo$ currents from the SQD to the left reservoir, as we depict in Fig.~\ref{fig:sv_currents_trajecs_both}a).

It is due to our feedback scheme B that the direction of the influx is reversed. The inflowing current, again, is split up and the $\spup$ electrons are transported to the right while the $\spdo$ electrons are released to the left; compare Fig.~\ref{fig:sv_currents_trajecs_both}c).

Interestingly, for scheme A and for the same parameters we find that the two transport channels decouple and the expectation values $\expval{\op{S}_i}$ decay; the averaged probabilities to find the SQD empty or occupied by an electron of spin $\sigma$ are equal. As a consequence, the $\spdo$ current through the left barrier vanishes, i.e., with Eq.~\eqref{eq:current_through_lbarrier} becomes $\expval{I_{L\spdo}} = (1-f_{L\spdo})/3 - \exp[\delta] f_{L\spdo}/3 = 0$. At the same time a constant net $\spup$ current from the left to the right lead builds up. This critical point is characterized by the ratio of $V/\delta = -2$. If the ratio is smaller than $-2$ a net outflux of $\spdo$ electrons to the left lead is generated while the $\spup$ current flows from left to right. For bias voltages $V > -2\delta$ $\spdo$ electrons are trapped in the SQD and only leave it after being flipped. In Fig.~\ref{fig:sv_currents_trajecs_both}e) we depicted the corresponding currents at and close to the critical parameters.

It is obvious that $V=0$ is the critical bias voltages for the nonfeedback and scheme B as well, where the two transport channels decouple. If one can prepare the LS to have a polarization of $\expval{\op{J}_z}$ the energies $\varepsilon_\sigma$ are constant and equal to the chemical potentials of the reservoirs. It is due to the thermal melting of the Fermi edges that transport is still possible, and therefore we get the net currents \eqref{eq:spincurrent_no_large_spin} evaluated for $B=0$ since the level splitting is compensated by the LS polarization.

We should also mention, that the opposite oscillating currents can as well be generated when the tunneling rates are chosen symmetrically ($\Gamma_{l\sigma} = \Gamma$), and therefore, spin filter behavior is possible in setups without polarized leads.

\section{\label{sec:conclusions}Conclusions}
In this work we have studied the complex dynamics of electronic transport through a SQD when interacting with a large external spin. We use semiclassical EOM for the large external spin together with a quantum master equation technique for the dynamics of the SQD. This method works well if we assume that the LS precession is taking place on a much slower time scale than the electronic tunneling. The treatment allows for the introduction of a Maxwell-demon-like feedback. This feedback uses occupation-dependent alternation of the transport Liouvillian and can generate currents even against a moderate bias voltage.

We restricted ourselves to studying the special case of an anisotropic exchange interaction and a polarized right transport lead, since this setup generates a variety of interesting dynamics, such as parametric oscillation of both electron spin and LS or complete polarization. We can calculate a number of fixed points analytically, for both finite and infinite-bias and under the influence of our demon's feedback.

Furthermore, we studied the effect of two different feedback schemes, which can work as an effective spin filter for small bias voltages, that ``sort'' electrons by its spin and generate oscillating currents in opposite directions.

For the infinite-bias regime both feedback schemes are identical. The numerical solutions of the highly nonlinear equations are very sensitive to the change of initial conditions and the parameters $(B,\Gamma,\delta)$ with respect to $\lambda$. 

On the one hand, the asymmetric transport setup as a consequence of the applied feedback can transfer periodic motions to quasiperiodic dynamics or even chaos, e.g., for magnetic fields close to $B_c$ and small tunneling rates $\Gamma \lessapprox \lambda$. 

On the other hand, we have found that for small magnetic fields $B \lessapprox 0.2 \lambda$ and tunneling rates $\Gamma \approx 0.1 \lambda$ quasiperiodic oscillations are changed to undamped periodic oscillations when applying feedback of moderate strengths ($\delta \approx 1$). 
\\
\\
As experimental realizations for the large external spin we think of magnetic impurities in semiconductor QDs or magnetic moments in single molecules (single molecular magnets). A possible realization where the mean-field approach is justified might be the hyperfine interaction with an ensemble of nuclear spins, where the number of spins is reasonably high. Although experiments suggest that it is possible to determine the spin state of electrons in a device it still is an open question as to how the spin-dependent tunneling rates can be modified conditionally in experiments.
 
On the other hand, the usage of conventional electronic circuits to control the 
tunneling setup seems also feasible. In experiments it is possible to measure single
electron tunneling events accurately and is possible to modify
tunneling rates on very fast time scales, by detuning gate voltages~\cite{Fricke2010} to access feedback parameters $\ln[\Gamma_{\mbs{filled}}/\Gamma_{\mbs{empty}}] = \delta \approx 1$, which lead to the mentioned feedback effect.


\begin{acknowledgments}
The authors would like to thank the DFG (Grant no. SFB 910 and Projects No. BR 1528/8-2 and No. BR 1528/9-1) for financial support.
\end{acknowledgments}


\begin{widetext}
\appendix
\renewcommand{\theequation}{\thesection.\arabic{equation}}
\section{\label{sec:derivation_master_equation}Derivation of the master equation}
In terms of a system bath theory the total Hamiltonian \eqref{eq:hamiltonian} consists of three parts, namely,
\begin{align}
 \op{H} &= \op{H}_{\mbs{S}} + \op{H}_{\mbs{B}} + \op{H}_{\mbs{SB}}\,,
\end{align}
where the system part $\op{H}_{\mbs{S}} = \op{H}_{\mbs{SQD}} + \op{H}_{\mbs{LS}} + \op{H}_{\mbs{int}}$ comprises the dynamics of the system, i.e., the SQD and the LS and the interaction between them. The electronic leads form the bath described by $\op{H}_{\mbs{B}} = \op{H}_{\mbs{leads}}$ and the coupling between system and bath degrees of freedom is denoted by the interaction Hamiltonian $\op{H}_{\mbs{SB}} = \op{H}_{\mbs{T}}$, which can be written in terms of system and bath coupling operators, which act on the respective Hilbert spaces only,
 \begin{align}
  \op{H}_{\mbs{T}} &= \sum_\alpha \op{A}_\alpha \otimes \op{B}_\alpha \,.
 \end{align}
 All Hamiltonians are required to be self-adjoint, as we consider physical observables. As $\op{H}_{\mbs{T}} = \op{H}_{\mbs{T}}^\dagger$ holds it is always possible to choose $\op{A}_\alpha, \op{B}_\alpha$ such that $\op{A}_\alpha = \dop{A}{\alpha}, \op{B}_\alpha = \dop{B}{\alpha}$.

 If the interaction $\op{H}_{\mbs{T}}$ is small, it is justified to apply perturbation theory. The von-Neumann equation describes the full evolution of the combined density matrix ($\hbar = 1$)
 \begin{align}
  \frac{\partial}{\partial t} \op{\rho} = -\imath \comm{\op{H}_{\mbs{S}} \otimes \mathds{1} + \mathds{1} \otimes \op{H}_{\mbs{B}} + \op{H}_{\mbs{T}}}{\op{\rho}} \,.
 \end{align}
 Its formal solution by unitary evolution, i.e. $\op{\rho}(t) = \e^{-\imath \op{H} t} \op{\rho}_0 \e^{+\imath \op{H} t}$ is impractical because $\op{H}$ involves too many degrees of freedom.
 Therefore, we switch to the interaction picture (operators denoted by bold symbols), such that the EOM for the density matrix reads
 \begin{align}
%
 \frac{\partial}{\partial t} \ip{\op{\rho}}(t) &= -\imath \comm{\ip{\op{H}_{\mbs{T}}}(t)}{\ip{\op{\rho}}(t)} \,, \label{eq:vNeumann_eq_interaction_picture}
\end{align}
where the interaction Hamiltonian in the interaction picture reads
\begin{align}
  \ip{\op{H}_{\mbs{T}}}(t) &= \e^{+\imath(\op{H}_{\mbs{S}} + \op{H}_{\mbs{B}})t} \op{H}_{\mbs{T}} \e^{-\imath(\op{H}_{\mbs{S}} + \op{H}_{\mbs{B}})t} = \sum_\alpha \e^{+\imath \op{H}_{\mbs{S}} t} \op{A}_\alpha \e^{-\imath \op{H}_{\mbs{S}} t} \otimes \e^{+\imath \op{H}_{\mbs{B}} t} \op{B}_\alpha \e^{-\imath \op{H}_{\mbs{B}} t} = \sum_\alpha \ip{\op{A}_\alpha}(t) \otimes \ip{\op{B}_\alpha}(t) \,.
 \end{align}
The formal integration of \eqref{eq:vNeumann_eq_interaction_picture} and re-inserting the result into the right-hand side of \eqref{eq:vNeumann_eq_interaction_picture} and then taking the partial trace leads to
\begin{align}
  \frac{\partial}{\partial t} \ip{\op{\rho}_{\mbs{S}}} &= -\imath \trb{\comm{\ip{\op{H}_{\mbs{T}}}(t)}{\rho_0}} - \int_0^t \trb{\comm{\ip{\op{H}_{\mbs{T}}}(t)}{\comm{\ip{\op{H}_{\mbs{T}}}(t')}{\ip{\op{\rho}}(t')}}dt'}\,. 
 \end{align}

Along the lines of Appendix A and B in our Ref.~\onlinecite{Schaller2008} we make the following simplifying assumptions.
(i) Initial factorization of the density matrix: $\ip{\op{\rho}}(0) = \ip{\op{\rho}_{\mbs{S}}} \otimes \ip{\op{\rho}_{\mbs{B}}^0} = \op{\rho}_{\mbs{S}}^0 \otimes \op{\rho}_{\mbs{B}}^0$.
(ii) Born approximation: If the bath is much larger than the system and the system-bath coupling, one can assume that the backaction of the system onto the bath is small; i.e., the bath density matrix is hardly changed from its initial state: $\ip{\op{\rho}}(t) = \ip{\op{\rho}_{\mbs{S}}}(t) \otimes \op{\rho}_{\mbs{B}}^0$.


 The Born approximation is equivalent to a second-order perturbation theory in the interaction Hamiltonian, i.e.,
 \begin{align}
  \frac{\partial}{\partial t} \ip{\op{\rho}_{\mbs{S}}} &= -\imath \trb{\comm{\ip{\op{H}_{\mbs{T}}}(t)}{\rho_0}} - \int_0^t \trb{\comm{\ip{\op{H}_{\mbs{T}}}(t)}{\comm{\ip{\op{H}_{\mbs{T}}}(t')}{\ip{\op{\rho}_{\mbs{S}}}(t') \otimes \op{\rho}_{\mbs{B}}^0}}dt'}\,. \label{eq:Born_equation}
 \end{align}

The Born equation \eqref{eq:Born_equation} further simplifies by noting that the single coupling operator expectation value vanishes $\tr{\ip{\op{B}_\alpha}(t) \op{\rho}_{\mbs{B}}^0}$. When evaluating the traces over the bath degrees of freedom, $\trb{\cdots}$, the bath-correlation functions are introduced,
\begin{align}
 C_{\alpha \beta}(t_1,t_2) &= \tr{\ip{\op{B}_\alpha}(t_1) \ip{\op{B}_\beta}(t_2)\op{\rho}_{\mbs{B}}^0} \,,
\end{align}
with the the following analytic properties. (i) Given that the bath density matrix is in its stationary state, i.e., $\comm{\op{H}_{\mbs{B}}}{\op{\rho}_{\mbs{B}}^0} = 0$, the bath correlation functions just depend on the difference of their time arguments, $C_{\alpha \beta}(t_1,t_2) = C_{\alpha \beta}(t_1 - t_2) = \tr{\e^{+\imath \op{H}_{\mbs{B}}(t_1-t_2)}\op{B}_\alpha \e^{-\imath \op{H}_{\mbs{B}}(t_1-t_2)} \op{B}_\beta \op{\rho}_{\mbs{B}}^0}$. (ii) Bath coupling operators are chosen to be self-adjoint $C_{\alpha \beta}(\tau) = C_{\beta \alpha}^*(-\tau)$. (iii) When bath has a dense spectrum bath correlation functions are typically strongly peaked around zero.

Thus, we get the following integro-differential equation, the right-hand side of which still depends on the density matrix at all previous times weighted by the bath correlation functions,
 \begin{align}
  \frac{\partial}{\partial t} \ip{\op{\rho}_{\mbs{S}}} &= -\sum_{\alpha \beta} \int_0^t dt' \Big[C_{\alpha \beta}(t,t') \comm{\ip{\op{A}_\alpha}(t)}{\ip{\op{A}_\beta}(t') \ip{\op{\rho}_{\mbs{S}}}(t')} + C_{\beta \alpha}(t',t) \comm{\ip{\op{\rho}_{\mbs{S}}}(t') \ip{\op{A}_\beta}(t') }{\ip{\op{A}_\alpha}(t)} \Big]\,.
 \end{align}

Now the Markov approximations are implemented in the usual way~\cite{Breuer2002}. In the Markov approximation we suppose that the variation of the reduced density
matrix is slower than the decay of the bath correlation function. That is, we replace $\ip{\op{\rho}_{\mbs{S}}}(t') \to \ip{\op{\rho}_{\mbs{S}}}(t)$ and substitute $\tau = t - t'$,
\begin{align}
  \frac{\partial}{\partial t} \ip{\op{\rho}_{\mbs{S}}} =& -\sum_{\alpha \beta} \int_0^\infty d\tau \Big[C_{\alpha \beta}(\tau) \comm{\ip{\op{A}_\alpha}(t)}{\ip{\op{A}_\beta}(t-\tau) \ip{\op{\rho}_{\mbs{S}}}(t)} \nnn 
 & + C_{\beta \alpha}(-\tau) \comm{\ip{\op{\rho}_{\mbs{S}}}(t) \ip{\op{A}_\beta}(t-\tau) }{\ip{\op{A}_\alpha}(t)} \Big]\,.
\label{eq:born_markov_me_fullform}
 \end{align}
In the last step we also applied the second Markov approximation, the extension of the integration boundary to ``$\infty$,'' which is justified if the bath correlation functions decay sufficiently fast.

We turn back to the Schr\"odinger picture and obtain
\begin{align}
  \frac{\partial}{\partial t} \op{\rho}_{\mbs{S}} = & -\imath \comm{\op{H}_{\mbs{S}}}{\op{\rho}_{\mbs{S}}(t)} - \sum_ {\alpha \beta} \int_0^\infty  d\tau \Big \{ C_{\alpha \beta}(\tau) \comm{\op{A}_\alpha}{\e^{-\imath \op{H}_{\mbs{S}} \tau} \op{A}_\beta \e^{+\imath \op{H}_{\mbs{S}} \tau} \op{\rho}_{\mbs{S}}(t)} \nnn
 & + C_{\beta \alpha}(-\tau) \comm{\op{\rho}_{\mbs{S}}(t) \e^{-\imath \op{H}_{\mbs{S}} \tau} \op{A}_\beta \e^{+\imath \op{H}_{\mbs{S}} \tau} }{\op{A}_\alpha} \Big \} \,, \label{eq:born_markov_me_fullform_schroedinger_picture}
\end{align}
which is time local, preserves trace and Hermiticity and has constant coefficients.

\subsection{\label{sec:mean-field_approximation}Mean field approximation}
We want to treat the interaction of electronic and LS in a semiclassical manner, which should be valid as long as the external spin is sufficiently large ($j \gg 1$) and we can neglect its fluctuations. Consequently, there is no decay of the LS due to dissipation and the length of the LS will be conserved on the microscopic level, $\left( \comm{\vec{\op{J}}^2}{\op{H}} = 0 \right)$. 

In a mean-field approach we rewrite the Hamiltonian describing the spin-spin interaction by substituting $\op{S}_i = \expval{\op{S}_i}+ \delta \op{S}_i$ and $\op{J}_i = \expval{\op{J}_i}+ \delta \op{J}_i$

\begin{align}
  \op{H}_{\mbs{int}}^{\mbs{MF}} &= \sum_{i} \lambda_i \left(\op{S}_i \expval{\op{J}_i} 
+ \expval{\op{S}_i} \op{J}_i - \expval{\op{S}_i} \expval{\op{J}_i} \right) \,. \label{eq:mean-field_interaction_hamiltonian}
\end{align}

The system Hamiltonian in the unitary part of \eqref{eq:born_markov_me_fullform_schroedinger_picture} is, accordingly, $\op{H}_{\mbs{S}} = \varepsilon\left(\op{n}_\spup + \op{n}_\spdo \right) + \frac{B}{2} \left(\op{n}_\spup - \op{n}_\spdo \right) + \op{H}_{\mbs{int}}^{\mbs{MF}} + \op{H}_{\mbs{LS}}$, where we find the following vanishing commutators $\comm{\expval{\op{S}_{i}} \op{J}_i}{\op{\rho}_{\mbs{S}}} = \comm{\expval{\op{S}_{i}} \expval{\op{J}_{i}} \mathds{1}}{\op{\rho}_{\mbs{S}}} = \comm{\op{H}_{\mbs{LS}}}{\op{\rho}_{\mbs{S}}} = 0$.

As a consequence we rewrite the system Hamiltonian as an effective Hamiltonian by comprising all terms contributing to electronic level shifts and all terms that participate in the spin flip processes,
\begin{align}
 \op{H}_{\mbs{S}}^{\mbs{eff}} &= \sum_\sigma \varepsilon_\sigma \op{n}_\sigma + \Lambda^* \op{S}_+ + \Lambda \op{S}_- \,, \mbox{with } \varepsilon_\sigma = \frac{B}{2} \pm \frac{\lambda_z}{2} \expval{\op{J}_z}_t \,, \Lambda = \frac{\lambda_x}{2} \expval{\op{J}_x}_t + \imath \frac{\lambda_y}{2} \expval{\op{J}_y}_t
\end{align}
Note, that the parameters are explicitly time dependent, although we assume the time scale set by $\varepsilon_\sigma$ and $\Lambda$ to be much slower than the timescale of the lead correlations $\tau_c$ and as well slower than the time scale of the electronic tunneling.

\subsection{System coupling operators}
In order to evaluate the master equation \eqref{eq:born_markov_me_fullform_schroedinger_picture} we have to identify system coupling operators
\begin{align}
 \op{A}_1 &= \op{A}_5 = \dop{A}{2} = \dop{A}{6} = \op{d}_\spup \,, \quad \op{A}_3 = \op{A}_7 = \dop{A}{4} = \dop{A}{8} = \op{d}_\spdo \,. \label{eq:system_coupling_ops}
\end{align}

In a next step we calculate the interaction picture for the system operators $\op{A}_\alpha$ by neglecting the spin flip terms in the effective Hamiltonian and use the free Hamiltonian instead, which reads
\begin{align}
 \op{H}^{\mbs{free}}_{\mbs{S}} &= \sum_\sigma \varepsilon_\sigma \dop{d}{\sigma} \op{d}_{\sigma}\,.
\end{align}
We replace $\op{H}_{\mbs{S}}^{\mbs{eff}} \to \op{H}_{\mbs{S}}^{\mbs{free}}$ in the integral in \eqref{eq:born_markov_me_fullform_schroedinger_picture}. Thus, we can calculate the terms $\e^{-\imath \op{H}_{\mbs{S}}^{\mbs{free}} \tau} \op{d}_\sigma^{(\dagger)} \e^{+\imath \op{H}_{\mbs{S}}^{\mbs{free}} \tau}$ by employing the Baker-Campbell-Hausdorff-formula, i.e.,
\begin{align}
 \e^X Y \e^{-X} &= \sum_{m=0}^\infty \frac{1}{m!} \comm{X}{Y}_m\,,
\end{align}
where $\comm{X}{Y}_m = \comm{X}{\comm{X}{Y}_{m-1}} \mbox{ and } \comm{X}{Y}_0 = Y$.
Thus we have to find the commutators $\comm{X}{Y}$ and use the fermionic (anti-)commutation relations
\begin{align}
 \comm{X}{Y} &= \comm{-\imath \op{H}^{\mbs{free}}_{\mbs{S}} \tau}{\op{d}_\sigma} = - \imath \sum_\mu \varepsilon_\mu \tau \comm{\dop{d}{\mu} \op{d}_\mu}{\op{d}_\sigma} = - \imath \sum_\mu \varepsilon_\mu \tau \left[\dop{d}{\mu} \anticomm{\op{d}_{\mu}}{\op{d}_\sigma} - \anticomm{\dop{d}{\mu}}{\op{d}_\mu} \op{d}_\sigma \right] \nnn 
& = \imath \sum_\mu \varepsilon_\mu \tau \delta_{\mu \sigma} \op{d}_\sigma = \imath \varepsilon_\sigma \tau \op{d}_\sigma \,. 
\end{align}
Applied recursively, we get
\begin{align}
 \e^{-\imath \op{H}_{\mbs{S}}^{\mbs{free}} \tau} \op{d}_\sigma \e^{+\imath \op{H}_{\mbs{S}}^{\mbs{free}} \tau} &= \e^{\imath \varepsilon_\sigma \tau} \op{d}_\sigma \,, \quad
 \e^{-\imath \op{H}_{\mbs{S}}^{\mbs{free}} \tau} \op{d}_\sigma^{\dagger} \e^{+\imath \op{H}_{\mbs{S}}^{\mbs{free}} \tau} = \e^{-\imath \varepsilon_\sigma \tau} \dop{d}{\sigma} \,.
\end{align}

\textit{Justification for the use of the free Hamiltonian}. For a convenient notation we split up the effective system Hamiltonian $\op{H}_{\mbs{S}}^{\mbs{eff}} = \op{H}_0 + \op{H}_1$, where $\op{H}_0 = \op{H}_{\mbs{S}}^{\mbs{free}}$ contains the two SQD levels and $\op{H}_1 = \op{H}_{\mbs{S}}^{\mbs{SF}}$ describes the spin flips due to the exchange interaction. The transformation of an arbitrary system operator $\op{A}$ to the interaction picture can be written in terms of the Schwinger-Dyson identity involving superoperators $\op{\call}_x \cdot = \comm{\op{H}_x}{\cdot}$
\begin{align}
 \e^{-\imath \op{\call}_{\mbs{S}} \tau} &= \e^{-\imath \left(\op{\call}_{0} + \op{\call}_1 \right) \tau} = \e^{-\imath \op{\call}_0 \tau} \sum_{n=0}^\infty (-\imath)^n \int_0^\tau d\tau_1 \int_0^{\tau_1} d\tau_2 \dots \int_0^{\tau_{n-1}} d\tau_n \ip{\op{\call}_1}(\tau_1) \ip{\op{\call}_1}(\tau_2) \cdots \ip{\op{\call}_1}(\tau_n)\,,
\end{align}
where for any operator $\op{A}$
\begin{align}
 \ip{\op{\call}_1}(\tau) \op{A} &= \e^{\imath \op{\call}_0 \tau} \op{\call}_1 \e^{-\imath \op{\call}_0 \tau} = \comm{\e^{\imath \op{H}_0 \tau} \op{H}_1 \e^{-\imath \op{H}_0 \tau}}{\op{A}} = \comm{\ip{\op{H}_1}(\tau)}{\op{A}}\,.
\end{align}
The interaction picture operator in terms of this expansion up to the first order reads
\begin{align}
  \e^{-\imath \op{H}_{\mbs{S}} \tau} \op{A} \e^{\imath \op{H}_{\mbs{S}} \tau} &= \e^{-\imath \left( \op{\call}_{0} + \op{\call}_{1} \right) \tau} \op{A} = \e^{-\imath \op{\call}_0 \tau} \op{A} - \imath \e^{-\imath \op{\call}_0 \tau} \int_0^\tau d\tau_1 \comm{\ip{\op{H}_1}(\tau_1)}{\op{A}} + \dots \,.
\end{align}
where we have to write the spin flip Hamiltonian $\op{H}_1$ in the interaction picture by applying, again, the Baker-Campbell-Hausdorff-formula,
\begin{align}
 \ip{\op{H}_1}(\tau) &= \e^{\imath \varepsilon_z \tau} \Lambda^* \op{S}_+ + \e^{-\imath \varepsilon_z \tau} \Lambda \op{S}_- \,,
\end{align}
which introduces the effective Zeeman-splitting $\varepsilon_z = \varepsilon_\spup - \varepsilon_\spdo = B + \lambda_z \expval{\op{J}_z}_t$.

The integrals in \eqref{eq:born_markov_me_fullform_schroedinger_picture} then read
\begin{align}
 (\ast) &= \int_0^\infty  d\tau C_{\alpha \beta}(\tau) \comm{\op{A}_\alpha}{\e^{-\imath \op{H}_{\mbs{S}} \tau} \op{A}_\beta \e^{+\imath \op{H}_{\mbs{S}} \tau} \op{\rho}_{\mbs{S}}(t)} \nnn 
 &= \int_0^\infty  d\tau C_{\alpha \beta}(\tau) \comm{\op{A}_\alpha}{\e^{-\imath \op{H}_0 \tau} \op{A}_\beta \e^{+\imath \op{H}_0 \tau} \op{\rho}_{\mbs{S}}(t)} + (\ast^{(1)}) \,, \nnn
(\ast^{(1)}) &= \int_0^\infty  d\tau C_{\alpha \beta}(\tau) \comm{\op{A}_\alpha}{-\imath \e^{-\imath \op{\call}_0 \tau} \int_0^\tau d\tau_1 \left\{ \e^{\imath \varepsilon_z \tau_1} \Lambda^* \comm{\op{S}_+}{\op{A}_\beta} + \e^{-\imath \varepsilon_z \tau_1} \Lambda \comm{\op{S}_-}{\op{A}_\beta}\right\} \op{\rho}_{\mbs{S}}(t)} \nnn
&= \int_0^\infty  d\tau C_{\alpha \beta}(\tau) \comm{\op{A}_\alpha}{-\imath \e^{-\imath \op{\call}_0 \tau} \left\{ \frac{\Lambda^*}{\varepsilon_z} \left(\e^{\imath \varepsilon_z \tau_1} - 1 \right) \comm{\op{S}_+}{\op{A}_\beta} + \frac{\Lambda}{\varepsilon_z} \left(1- \e^{-\imath \varepsilon_z \tau_1}\right) \comm{\op{S}_-}{\op{A}_\beta} \right \} \op{\rho}_{\mbs{S}}(t)} \,. \label{eq:first_order_correction_dissipator}
\end{align}

Here we have to consider two different time scales: the leads-correlation time scale $\tau_c$ that is in general much faster than the time scale set by the Zeeman splitting $\varepsilon_z$, i.e., $\varepsilon_z \tau_c \ll 1$. We can thus perform an expansion in the parameter $\varepsilon_z \tau$: $\frac{\Lambda}{\varepsilon_z}  \left(1 - \e^{-\imath \varepsilon_z \tau_1} \right) \approx - \imath \Lambda \tau$. As a consequence, it is justified to neglect the first-order corrections, $(\ast^{(1)})$ in \eqref{eq:first_order_correction_dissipator}, if the bath correlation time $\tau_c$ is sufficiently short compared to the time scale of the spin flip dynamics, i.e. $\Lambda \tau_c \ll 1$. Note, that Schuetz \textit{et.al.} have provided a calculation for higher order corrections in a similar setup which suggests that the $n$th order correction scales with $(\Lambda \tau_c)^n$~\cite{Schuetz2012}.

\subsection{Transition rates}
In order to calculate the transition rates we first identify the bath coupling operators in $\op{H}_{\mbs{T}}$,
\begin{align}
 \op{B}_1 &= \sum_k \gamma_{k R \spup}^* \dop{c}{kR\spup}\,, \quad \op{B}_2 = \sum_k \gamma_{k R \spup} \op{c}_{kR\spup}\,, \quad \op{B}_3 = \sum_k \gamma_{k R \spdo}^* \dop{c}{kR\spdo}\,, \quad \op{B}_4 = \sum_k \gamma_{k R \spdo} \op{c}_{kR\spdo}\,, \nnn
 \op{B}_5 &= \sum_k \gamma_{k L \spup}^* \dop{c}{kL\spup}\,, \quad \op{B}_6 = \sum_k \gamma_{k L \spup} \op{c}_{kL\spup}\,, \quad \op{B}_7 = \sum_k \gamma_{k L \spdo}^* \dop{c}{kL\spdo}\,, \quad \op{B}_8 = \sum_k t_{k L \spdo} \op{c}_{kL\spdo}\,. 
\end{align}

Next, we define the half-sided Fourier transformation of the bath-correlation functions and its splitting into Hermitian and anti-Hermitian parts
\begin{align}
 \Gamma_{\alpha \beta}^\pm(\omega) &= \int_{-\infty}^{\infty} d\tau \Theta(\pm \tau) C_{\alpha \beta}(\tau) \e^{\imath \omega \tau} \label{eq:half-sided_fourier_trafo}\,, \quad
 \Gamma_{\alpha \beta}^\pm(\omega) = \frac{1}{2} \gamma_{\alpha \beta}(\omega) \pm \frac{1}{2} \sigma_{\alpha \beta}(\omega) \,. 
\end{align}

With the identified system and bath operators we can exemplarily calculate the dissipator with respect to tunneling between the right lead and the $\spup$ state
\begin{align}
 (\star) = & \int_0^\infty d\tau C_{12}(\tau) \e^{-\imath \varepsilon_\spup \tau} \comm{\op{d}_\spup}{\dop{d}{\spup} \op{\rho}_{\mbs{S}}(t)} + \int_0^\infty d\tau C_{21}(-\tau) \e^{-\imath \varepsilon_\spup \tau} \comm{\dop{d}{\spup} \op{\rho}_{\mbs{S}}(t)}{\op{d}_\spup} \nnn
& + \int_0^\infty d\tau C_{21}(\tau) \e^{\imath \varepsilon_\spup \tau} \comm{\dop{d}{\spup}}{\op{d}_{\spup} \op{\rho}_{\mbs{S}}(t)} + \int_0^\infty d\tau C_{12}(-\tau) \e^{\imath \varepsilon_\spup \tau} \comm{\op{d}_{\spup} \op{\rho}_{\mbs{S}}(t)}{\dop{d}{\spup}}\,,
\end{align}
which reads with the definition for the half-sided Fourier transformations \eqref{eq:half-sided_fourier_trafo}
\begin{align}
 (\star) & =  \Gamma_{12}^+(-\varepsilon_\spup) \comm{\op{d}_\spup}{\dop{d}{\spup} \op{\rho}_{\mbs{S}}(t)} 
  + \Gamma_{12}^-(-\varepsilon_\spup) \comm{\op{d}_{\spup} \op{\rho}_{\mbs{S}}(t)}{\dop{d}{\spup}} + \Gamma_{21}^+(\varepsilon_\spup) \comm{\dop{d}{\spup}}{\op{d}_{\spup} \op{\rho}_{\mbs{S}}(t)} 
  + \Gamma_{21}^-(\varepsilon_\spup) \comm{\dop{d}{\spup} \op{\rho}_{\mbs{S}}(t)}{\op{d}_\spup} \nnn
%
%
& = \frac{1}{2} \gamma_{12}(-\varepsilon_\spup) \left(\anticomm{\op{d}_\spup \dop{d}{\spup}}{\op{\rho}_{\mbs{S}}} - 2 \dop{d}{\spup} \op{\rho}_{\mbs{S}}(t) \op{d}_\spup \right) + \frac{1}{2} \gamma_{21}(\varepsilon_\spup) \left(\anticomm{\dop{d}{\spup} \op{d}_{\spup}}{\op{\rho}_{\mbs{S}}} - 2 \op{d}_{\spup} \op{\rho}_{\mbs{S}}(t) \dop{d}{\spup} \right) \nnn
& \quad + \frac{1}{2} \left(\sigma_{21}(\varepsilon_\spup) - \sigma_{12}(-\varepsilon_\spup) \right) \comm{\dop{d}{\spup} \op{d}_\spup}{\op{\rho}_{\mbs{S}}(t)} \,.
\end{align}

Subsequently, we find the following non-vanishing coefficients (the even Fourier transformations of the bath correlation functions):
\begin{align}
 \gamma_{12}(\omega) &= \Gamma_{R\spup}(-\omega) f_R(-\omega)\,, \quad \quad \gamma_{21}(\omega) = \Gamma_{R\spup}(\omega) (1 - f_R(\omega))\,, \nnn
 \gamma_{34}(\omega) &= \Gamma_{R\spdo}(-\omega) f_R(-\omega)\,, \quad \quad \gamma_{43}(\omega) = \Gamma_{R\spdo}(\omega) (1 - f_R(\omega))\,, \nnn
 \gamma_{56}(\omega) &= \Gamma_{L\spup}(-\omega) f_L(-\omega)\,, \quad \quad \gamma_{65}(\omega) = \Gamma_{L\spup}(\omega) (1 - f_L(\omega))\,, \nnn
 \gamma_{78}(\omega) &= \Gamma_{L\spdo}(-\omega) f_L(-\omega)\,, \quad \quad \gamma_{87}(\omega) = \Gamma_{L\spdo}(\omega) (1 - f_L(\omega))\,. \label{eq:even_fourier_transforms}
\end{align}

Here we defined the the Fermi function of the reservoir $l$ as 
\begin{align}
 f_{l} (\omega) &= \frac{1}{\e^{\beta_l(\omega - \mu_l)} + 1}\,,
\end{align}
evaluated at the respective transition frequencies $\omega$ and the energy-dependent tunneling rates between the lead $l$ and the spin-$\sigma$ level of the QD,
\begin{align}
 \Gamma_{l\sigma}(\omega) &= 2\pi \sum_k |\gamma_{kl\sigma}|^2 \delta(\varepsilon_{kl\sigma} - \omega)\,,
\end{align}
where $\beta_l = (k_B T_l)^{-1}$ is the inverse temperature of the leads.

Additionally, the odd Fourier transforms of the bath correlation functions, called Lamb-shift terms, are calculated from the even Fourier transforms by using a Cauchy principal value integral,
\begin{align}
 \sigma_{\alpha \beta} (\omega) &= \frac{\imath}{\pi} \calp \int_{-\infty}^\infty \frac{\gamma_{\alpha \beta}(\Omega)}{\omega - \Omega} d\Omega \,. 
\end{align}
We show, straightforwardly, that the Lamb-shift contributions vanish,
\begin{align}
 \sigma_{21}(\omega) - \sigma_{12}(-\omega) &= \frac{\imath}{\pi} \calp \int_{-\infty}^\infty \frac{\Gamma_{R\spup} \left[f_R(\Omega) + 1 - f_R(\Omega) \right] }{ \omega - \Omega} d\Omega \nnn
&= \frac{\imath}{\pi} \calp \int_{-\infty}^\infty \frac{\Gamma_{R\spup}}{\omega - \Omega} d\Omega = \frac{\imath}{\pi} \lim_{\omega' \to \infty} \calp \int_{-\omega'}^{\omega'} \frac{\Gamma_{R\spup}}{\omega - \Omega} d\Omega = \frac{\imath}{\pi} \Gamma_{R\spup} \lim_{\omega' \to \infty} \left(\frac{|\omega-\omega'|}{|\omega + \omega'|} \right) = 0 \,,
\end{align}
where we set energy-independent tunneling rates; i.e., the density of states around the transition frequencies is assumed to be flat (wideband limit).

The final master equation has Lindblad form, i.e., preserves trace, Hermiticity, and the positivity of the reduced density matrix,
\begin{align}
 \frac{\partial}{\partial t} \op{\rho}_{\mbs{S}}(t) &= \op{\call}[\op{\rho}_{\mbs{S}}(t)] = -\imath \comm{\op{H}_{\mbs{S}}}{\op{\rho}_{\mbs{S}}(t)} + \sum_{k=1}^{N^2 - 1} \gamma_k \left(\op{A}_k \op{\rho}_{\mbs{S}}(t) \dop{A}{k} - \frac{1}{2} \anticomm{\dop{A}{k} \op{A}_k}{ \op{\rho}_{\mbs{S}}(t)} \right)\,,  \label{eq:me_with_dissipators_full_form}
\end{align}
where $N=3$ is the dimension of the SQD Hilbert space and we used \eqref{eq:system_coupling_ops}. The non-negative eigenvalues of the generator $\op{\call}$ are the rates 
\begin{align}
 \gamma_1 &= \gamma_{21}(\varepsilon_\spup) \,,\quad \gamma_2 = \gamma_{12}(-\varepsilon_\spup) \,, \quad \gamma_3 = \gamma_{43}(\varepsilon_\spdo) \,,\quad \gamma_4 = \gamma_{34}(-\varepsilon_\spdo) \,, \nnn
 \gamma_5 &= \gamma_{65}(\varepsilon_\spup) \,,\quad \gamma_6 = \gamma_{56}(-\varepsilon_\spup) \,, \quad \gamma_7 = \gamma_{87}(\varepsilon_\spdo) \,,\quad \gamma_8 = \gamma_{78}(-\varepsilon_\spdo) \,.
\end{align}

\subsection{Complete Liouvillian}
We now take matrix elements in an arbitrary basis, $\rho_{ij} = \bra{i} \op{\rho}_{\mbs{S}} \ket{j}$, such that

\begin{align}
 \dot{\rho}_{ij} &= -\imath \bra{i}\comm{\op{H}_{\mbs{S}}}{\op{\rho}_{\mbs{S}}} \ket{j}
 + \sum_{k=1}^{8} \gamma_k \left( \bra{i} \op{A}_k \op{\rho}_{\mbs{S}}(t) \dop{A}{k} \ket{j} - \frac{1}{2} \bra{i} \anticomm{\dop{A}{k} \op{A}_k}{ \op{\rho}_{\mbs{S}}(t)} \ket{j} \right) \,,
\end{align}
and obtain in the local basis ($i,j \in {0,\spdo,\spup}$). Taking the expectation values for the relevant observables $\frac{d}{dt} \expval{\op{O}} = \trs{\op{O} \op{\call}[\op{\rho}_{\mbs{S}}(t)]}$ and writing the vector
\begin{align}
 \vec{\rho} &= \left[\expval{\rho_{00}}, \expval{\op{n}_{\spdo}}, \expval{\op{n}_{\spup}}, \expval{\op{S}_+},  \expval{\op{S}_-} \right]^T\,,
\end{align}

we finally obtain the rate equations describing the dynamics of the interesting observables we are interested in:
\begin{align}
 \frac{\partial}{\partial t} \vec{\rho} &= \call \vec{\rho}\,.
\end{align}
The Liouvillian is represented by the rate matrix $\call = \sum_l \call^{(l)}$, which reads
\begin{align}
 \call^{(l)} &= \parmat{
- \sum_{\sigma} \Gamma_{l\sigma} f_{l\sigma} &  \Gamma_{l\spdo} \overline{f_{l\spdo}} & \Gamma_{l\spup}
   \overline{f_{l\spup}} & 0 & 0 \\
  \Gamma_{l\spdo} f_{l\spdo} & -  \Gamma_{l\spdo} \overline{f_{l\spdo}} & 0 & \frac{\imath}{2} \Lambda^* & -\frac{\imath}{2} \Lambda \\
  \Gamma_{l\spup} f_{l\spup} & 0 & - \Gamma_{l\spup} \overline{f_{l\spup}} & -\frac{\imath}{2} \Lambda^* & \frac{\imath}{2} \Lambda \\
 0 & \frac{\imath}{2} \Lambda & -\frac{\imath}{2} \Lambda & \frac{\imath}{2} \varepsilon_z - \frac{1}{2} \sum_{\sigma} \Gamma_{l\sigma} \overline{f_{l\sigma}}  & 0 \\
 0 & -\frac{\imath}{2} \Lambda^* & \frac{\imath}{2} \Lambda^* & 0 & -\frac{\imath}{2} \varepsilon_z - \frac{1}{2} \sum_{\sigma} \Gamma_{l\sigma} \overline{f_{l\sigma}}
}\,. \label{eq:full_Liouvillian}
\end{align}
The Fermi functions are evaluated at the transition frequencies $\varepsilon_\sigma$; thus, $f_{l\sigma} = f_l(\varepsilon_\sigma)$ and we define $\overline{f_{l\sigma}} = \left[ 1-f_{l\sigma} \right]$.

\section{\label{sec:derivation_effective_model}Simplified model for limit oscillation around $\vec{J}_z = -B/\lambda$}

For this model $\expval{\op{J}_z}$ is fixed and we transform the equations to a rotation-invariant frame assuming by using the rotation matrix around the $z$ axis to obtain stationary $\vec{\tilde{J}}$. Note that the effective rotation frequency $B_{\mbs{eff}}$ is not $B$, since the backaction of the electronic system on the precessing LS is changing this frequency:

\begin{align}
 \expval{\vec{\op{S}}} &= \e^{-\Gamma t} \mat{R}_S(t) \expval{\vec{\tilde{S}}}\,, \quad \quad \mat{R}_S(t) = \begin{pmatrix}
                                                                                      \cos(B_{\mbs{eff}} t) & -\sin(B_{\mbs{eff}} t)& 0 & 0\\
										      \sin(B_{\mbs{eff}} t) & \cos(B_{\mbs{eff}} t) & 0 & 0\\
											  0 & 0 & 1 & 0 \\ 
											  0 & 0 & 0 & 1 
                                                                                    \end{pmatrix} \,, \nnn
 \expval{\vec{\op{J}}} &=  \mat{R}_J(t) \expval{\vec{\tilde{J}}}\,, \quad \quad \mat{R}_J(t) = \begin{pmatrix}
                                                                                      \cos(B_{\mbs{eff}} t) & -\sin(B_{\mbs{eff}} t)& 0 \\
										      \sin(B_{\mbs{eff}} t) & \cos(B_{\mbs{eff}} t) & 0 \\
											  0 & 0 & 1
                                                                                    \end{pmatrix}\,.
\end{align}

Using the inverse transformations $\expval{\vec{\tilde{S}}} = \e^{\Gamma t} \mat{R}_S^{-1} \expval{\vec{\op{S}}}$ and $\expval{\vec{\tilde{J}}} = \mat{R}_J^{-1} \expval{\vec{\op{J}}}$ we can calculate the eom for the spin components in the rotation invariant frame:

\begin{align}
\frac{d}{dt} \expval{\vec{\tilde{S}}} 
 &=  \e^{\Gamma t} \mat{R}_S^{-1} \frac{d \expval{\vec{\op{S}}}}{dt} + \frac{d\mat{R}_S^{-1}}{dt} \mat{R}_S \expval{\vec{\tilde{S}}} + \Gamma \mat{R}_S^{-1} \mat{R}_S \expval{\vec{\tilde{S}}} \nnn
\frac{d}{dt} \expval{\vec{\tilde{J}}} 
&= \mat{R}_J^{-1} \frac{d \expval{\vec{\op{J}}}}{dt} + \frac{d\mat{R}_J^{-1}}{dt} \mat{R}_J \expval{\vec{\tilde{J}}} \,.
\end{align}
Inserting Eqs.~\eqref{eq:eom_sqd} and \eqref{eq:eom_largespin_expanded} and applying the spin-valve feedback scheme, anisotropic coupling, and the infinite-bias setup with tunneling rates $\Gamma_{R\spdo} = 0, \Gamma_{L\spup} = \Gamma_{R\spup} = \Gamma_{L\spdo} = \Gamma$, we obtain
\begin{align}
 \frac{d}{dt} \expval{\tilde{S}_x} &= -\lambda \left(\expval{\tilde{J}_x} \cos(B_{\mbs{eff}} t) - \expval{\tilde{J}_y} \sin(B_{\mbs{eff}} t) \right) \sin (B_{\mbs{eff}} t) \expval{\tilde{S}_z} - \lambda \expval{\tilde{J}_z} \expval{\tilde{S}_y} +\frac{\Gamma}{2} \expval{\tilde{S}_x} \nnn
 \frac{d}{dt} \expval{\tilde{S}_y} &= -\lambda \left(\expval{\tilde{J}_x} \cos(B_{\mbs{eff}} t) - \expval{\tilde{J}_y} \sin(B_{\mbs{eff}} t) \right) \cos (B_{\mbs{eff}} t) \expval{\tilde{S}_z} + \lambda \expval{\tilde{J}_z} \expval{\tilde{S}_x} +\frac{\Gamma}{2} \expval{\tilde{S}_y} \nnn
 \frac{d}{dt} \expval{\tilde{n}_\spup} &= \lambda\left[\left(\expval{\tilde{J}_x} \cos(B_{\mbs{eff}} t) - \expval{\tilde{J}_y} \sin(B_{\mbs{eff}} t) \right) \left(\expval{\tilde{S}_x} \sin(B_{\mbs{eff}} t) + \expval{\tilde{S}_y} \cos(B_{\mbs{eff}} t) \right)\right] \nnn
&  +\Gamma \left(1 - 2 \e^{\delta} \right) \expval{\tilde{n}_\spup} - \e^{\delta} \Gamma \expval{\tilde{n}_\spdo} + \e^{\delta - \Gamma t} \Gamma \nnn
 \frac{d}{dt} \expval{\tilde{n}_\spdo} &= - \lambda\left[\left(\expval{\tilde{J}_x} \cos(B_{\mbs{eff}} t) - \expval{\tilde{J}_y} \sin(B_{\mbs{eff}} t) \right) \left(\expval{\tilde{S}_x} \sin(B_{\mbs{eff}} t) + \expval{\tilde{S}_y} \cos(B_{\mbs{eff}} t) \right)\right] \nnn
&  +\Gamma \left(1 - \e^{\delta} \right) \expval{\tilde{n}_\spdo} - \e^{\delta} \Gamma \expval{\tilde{n}_\spup} + \e^{\delta - \Gamma t} \Gamma \nnn
\frac{d}{dt} \expval{\tilde{J}_x} &= - \lambda \e^{-\Gamma t} \left[\left(\expval{\tilde{S}_x} \cos(B_{\mbs{eff}} t) - \expval{\tilde{S}_y} \sin(B_{\mbs{eff}} t) \right) \sin (B_{\mbs{eff}} t) \expval{\tilde{J}_z} + \expval{\tilde{S}_z} \expval{\tilde{J}_y} \right] \nnn
\frac{d}{dt} \expval{\tilde{J}_y} &= \lambda \e^{-\Gamma t} \left[-\left(\expval{\tilde{S}_x} \cos(B_{\mbs{eff}} t) - \expval{\tilde{S}_y} \sin(B_{\mbs{eff}} t) \right) \cos (B_{\mbs{eff}} t) \expval{\tilde{J}_z} + \expval{\tilde{S}_z} \expval{\tilde{J}_x} \right] \nnn
\frac{d}{dt} \expval{\tilde{J}_z} &= \lambda \e^{-\Gamma t} \left[\left(\expval{\tilde{S}_x} \cos(B_{\mbs{eff}} t) - \expval{\tilde{S}_y} \sin(B_{\mbs{eff}} t) \right) \left(\expval{\tilde{J}_x} \sin(B_{\mbs{eff}} t) + \expval{\tilde{J}_y} \cos(B_{\mbs{eff}} t) \right)\right]
\end{align}
In the long-time limit the equations for $\vec{\tilde{J}}$ are stationary; thus, $\tilde{J}_i$ are constants in the equations for $\vec{\tilde{S}}$, and thus the number of equations is reduced in this effective model. From the numerical solutions we know that the LS's $z$ component assumes a characteristic value that is related two the fixed points $\calp_{\mbs{IB}}^{\pm y,\pm}$, namely, $\expval{\op{J}_z} = \expval{\tilde{J}_z} = -B/\lambda$, and since the LS is precessing almost unperturbed we can further assume $\expval{\tilde{J}_x} = \expval{\tilde{J}_y} = \frac{1}{\sqrt{2}} \sqrt{j^2-\frac{B^2}{\lambda^2}}$. The reduced set of equations for the effective model is obtained by applying the inverse transformation. We get 
\begin{align}
\frac{d}{dt} \expval{\vec{\op{S}}} &= \e^{-\Gamma t} \mat{R}_S \frac{\expval{\vec{\tilde{S}}}}{dt} + \frac{d\mat{R}_S}{dt} \mat{R}_S^{-1} \expval{\vec{\op{S}}} - \Gamma \expval{\vec{\op{S}}} \,,
\end{align}
which leads to \eqref{eq:effective_model}.

\section{\label{sec:app_finitebias_fixedpoints}Finite bias fixed points and stationary states}
The $\expval{\op{S}_z^*}$-components of the fixed points $\calp_{\mbs{FB}}^\pm$ read for scheme A and B
\begin{align}
 \calb_{4 \pm}^{\mbs{A}} &= \frac{\e^{\tilde{V}/2} \left(-\e^{\delta+\tilde{\varepsilon}_z^\pm}+\e^{2 \delta} + 1 \right) - \e^{3 \delta+\tilde{\varepsilon}_z^\pm+\frac{3 \tilde{V}}{2}} - \left(\e^{2
   \delta+\tilde{\varepsilon}_z^\pm}-\e^{\delta}+\e^{\tilde{\varepsilon}_z^\pm} \right) \e^{\delta+\frac{\tilde{\varepsilon}_z^\pm}{2}+\tilde{V}}+\e^{\tilde{\varepsilon}_z^\pm/2}}
{2 \left(\left(\e^{2 \delta+\tilde{\varepsilon}_z^\pm}+2 \e^{\delta}+\e^{\tilde{\varepsilon}_z^\pm}\right) \e^{\delta+\frac{\tilde{\varepsilon}_z^\pm}{2} + \tilde{V}} + \e^{3 \delta + \tilde{\varepsilon}_z^\pm + \frac{3 \tilde{V}}{2}} + \e^{\tilde{V}/2} \left(\left(\e^{\delta} + \e^{2 \delta} + 1 \right) \e^{\tilde{\varepsilon}_z^\pm} + \e^{2
   \delta} + 1\right) + 2 \e^{\tilde{\varepsilon}_z^\pm/2}\right)} \,,\nnn
\calb_{4 \pm}^{\mbs{B}} &= \frac{1}{2} \left(\frac{\left(\e^{2 \delta}+1\right) \left(\e^{\tilde{\varepsilon}_z^\pm}+2\right) \e^{\tilde{V}/2}+3 \e^{2
   \delta+\frac{\tilde{\varepsilon}_z^\pm}{2}+\tilde{V}}+3 \e^{\tilde{\varepsilon}_z^\pm/2}}
{\left(\e^{2 \delta}+1\right) \e^{\frac{3}{2} \tilde{\varepsilon}_z^\pm + \tilde{V}}+\e^{2
   \delta+\tilde{\varepsilon}_z^\pm+\frac{\tilde{V}}{2}}+2 \e^{2 \delta+\frac{\tilde{\varepsilon}_z^\pm}{2}+\tilde{V}}+\e^{2 \delta+\tilde{\varepsilon}_z^\pm+\frac{3 \tilde{V}}{2}}+\e^{2
   \delta+\frac{\tilde{V}}{2}}+2 \e^{\tilde{\varepsilon}_z^\pm+\frac{\tilde{V}}{2}}+2 \e^{\tilde{\varepsilon}_z^\pm/2}+\e^{\tilde{V}/2}}-1 \right) \,,
\end{align}
with the parameters
\begin{align}
 \tilde{\varepsilon}_z^\pm &\equiv \beta (B \pm \lambda j) \,, \tilde{V} \equiv \beta V\,. \nn
\end{align}

If one of the fixed points $\calp_{\mbs{FB}}^\pm$ is reached the exchange interaction becomes ineffective and the two spin-current channels decouple. Due to the tunneling setup $\Gamma_{L\spup} = \Gamma_{L\spdo} = \Gamma_{R\spup} = \Gamma, \Gamma_{R\spdo} = 0$, there is no net $\spdo$ current ($\expval{I_{L\spdo}^{\pm}} = - \expval{I_{R\spdo}^{\pm}} = 0$) and for $\spup$ current we obtain

 $\expval{I_{R\spup}^{\pm}} = - \expval{I_{L\spup}^{\pm}} = \expval{I_{\spup}^{\pm}}$,

which read with respect to the two feedback schemes
\begin{align}
\frac{\expval{I_{\spup}^{+{\mbs{A}}}}}{e\Gamma} &= \frac{ \e^{\frac{\tilde{\varepsilon}_z^{+}}{2}-\delta} \left(\e^{4 \delta+\tilde{V}}-1\right)}{\e^{\delta+\tilde{\varepsilon}_z^{+}+\frac{\tilde{V}}{2}}+\e^{2
   \delta+\tilde{\varepsilon}_z^{+}+\frac{\tilde{V}}{2}}+\e^{\delta+\frac{3 \tilde{\varepsilon}_z^{+}}{2}+\tilde{V}}+\e^{3 \delta+\frac{3 \tilde{\varepsilon}_z^{+}}{2}+\tilde{V}}+\e^{3
   \delta+\tilde{\varepsilon}_z^{+}+\frac{3 \tilde{V}}{2}}+2 \e^{2 \delta+\frac{1}{2} (\tilde{\varepsilon}_z^{+}+2 \tilde{V})}+\e^{2
   \delta+\frac{\tilde{V}}{2}}+\e^{\tilde{\varepsilon}_z^{+}+\frac{\tilde{V}}{2}}+2 \e^{\tilde{\varepsilon}_z^{+}/2}+\e^{\tilde{V}/2}}\,, \nnn
\frac{\expval{I_{\spup}^{-{\mbs{A}}}}}{e\Gamma} 
%
%
&= \frac{\e^{\tilde{a}-\delta} \left(\e^{4 \delta+\tilde{V}}-1\right)}
{\e^{2
   \delta} \left(\e^{\frac{1}{2} (\tilde{d}+V)}+2
   \e^{\tilde{a}+V}+\e^{\frac{\tilde{b}+\tilde{V}}{2}}\right)+\e^{3 \delta}
   \left(\e^{\frac{1}{2} (\tilde{d}+3
   V)}+\e^{\tilde{c}+\tilde{V}}\right)+\e^{\frac{1}{2}
   (\tilde{d}+V)+\delta}+\e^{\frac{1}{2} (\tilde{d}+V)}+2
   \e^{\tilde{a}}+\e^{\frac{\tilde{b}+\tilde{V}}{2}}+\e^{\tilde{c}+\delta+\tilde{V}}} \,,\nnn
\frac{\expval{I_{\spup}^{+{\mbs{B}}}}}{e \Gamma} &= \frac{ \e^{\frac{\tilde{\varepsilon}_z^{+}}{2}-\delta} \left(\e^{4 \delta+\tilde{V}}-1\right)}{\e^{2 \delta}
   \left(\e^{\tilde{\varepsilon}_z^{+}+\frac{\tilde{V}}{2}}+\e^{\tilde{\varepsilon}_z^{+}+\frac{3 \tilde{V}}{2}}+2 \e^{\frac{1}{2} (\tilde{\varepsilon}_z^{+}+2 \tilde{V})}+\e^{\frac{1}{2} (3 \tilde{\varepsilon}_z^{+}+2
   \tilde{V})}+\e^{\tilde{V}/2}\right)+2 \e^{\tilde{\varepsilon}_z^{+}+\frac{\tilde{V}}{2}}+\e^{\frac{1}{2} (3 \tilde{\varepsilon}_z^{+}+2 \tilde{V})}+2 \e^{\tilde{\varepsilon}_z^{+}/2}+\e^{\tilde{V}/2}} \,, \nnn
\frac{\expval{I_{\spup}^{-{\mbs{B}}}}}{e \Gamma} 
%
&= \frac{\e^{\tilde{a}-\delta} \left(\e^{4
   \delta+\tilde{V}}-1\right)}
{\e^{2 \delta} \left(\e^{\frac{1}{2}
   (\tilde{a}+\tilde{c}+\tilde{V})}+\e^{\frac{1}{2} (\tilde{a}+\tilde{c}+3
   \tilde{V})}+2
   \e^{\tilde{a}+\tilde{V}}+\e^{\frac{\tilde{b}+\tilde{V}}{2}}+\e^{\tilde{c}+\tilde{V}}\right)
   +2 \e^{\frac{1}{2} (\tilde{a}+\tilde{c}+\tilde{V})}+2
   \e^{\tilde{a}}+\e^{\frac{\tilde{b}+\tilde{V}}{2}}+\e^{\tilde{c}+\tilde{V}}}\,,
\end{align}
where the parameters 
$
 \tilde{a} \equiv \beta (B/2 + \lambda j) \,, \tilde{b} \equiv 3 \beta \lambda j \,, \tilde{c} \equiv 3 \beta B/2 \,, \tilde{d} \equiv \tilde{a}+\tilde{c}
$
have been introduced.

If the system has developed towards the fixed points $\calp_{\mbs{FB}}^{\pm y, \pm}$, we only need to consider the spin-dependent currents at the left lead since the currents are balanced ($\expval{I_{L\spup}^{y}} + \expval{I_{L\spdo}^{y}}  = - \expval{I_{R\spup}^{y}}$). We obtain

\begin{align}
 \frac{\expval{I_{L\spdo}^{y{\mbs{A}}}}}{e \Gamma} C_{\mbs{A}} &= \left(\e^{2 \delta+\frac{\tilde{V}}{2}}+1\right) \left((2 B - \lambda) \e^{\delta+\frac{\tilde{V}}{2}}+4 B + \lambda \right) \,,\quad 
C_{\mbs{A}} = \lambda \left(\e^{\tilde{V}/2}+1\right) \left(5 \e^{2 \delta+\frac{\tilde{V}}{2}}+\e^{\delta}+3\right) \,,\nnn
 \frac{\expval{I_{L\spup}^{y{\mbs{A}}}}}{e \Gamma} C_{\mbs{A}} &= \e^{-\delta} \left((2 B-\lambda) \e^{4 \delta+\tilde{V}}+\e^{2 \delta+\frac{\tilde{V}}{2}} \left(-\e^{\delta} (2
   B + \lambda) - 2 B + \lambda \right)-2 B \left(\e^{\delta}+1\right) + \lambda \right) \,,\nnn
 \frac{\expval{I_{L\spdo}^{y{\mbs{B}}}}}{e \Gamma} C_{\mbs{B}} &= \e^{\delta} \left(\e^{2 \delta+\frac{\tilde{V}}{2}}+1\right) \left(\e^{\tilde{V}/2} (2 B - \lambda) + 4 B + \lambda \right) \,,\quad
C_{\mbs{B}} = \lambda
   \left(\e^{\tilde{V}/2}+1\right) \left(\e^{2 \delta} \left(5 \e^{\tilde{V}/2}+1\right)+3\right)\,, \nnn
 \frac{\expval{I_{L\spup}^{y{\mbs{B}}}}}{e \Gamma} C_{\mbs{B}} &= \e^{-\delta} \left((2 B-\lambda) \e^{4 \delta+\tilde{V}}- 2 \e^{3 \delta+\frac{\tilde{V}}{2}} (2 B \cosh
   (\delta) + \lambda \sinh (\delta))-2 B \left(\e^{2 \delta}+1\right) + \lambda \right)\,. 
\end{align}

\end{widetext}

\bibliography{literature}
\end{document}